\documentclass[journal,comsoc]{IEEEtran}

\usepackage{latexsym}
\usepackage{epsfig}
\usepackage{color}
\usepackage{comment}
\usepackage[acronym]{glossaries}
\newacronym{tbps}{Tbps}{terabit-per-second}
\newacronym{thz}{THz}{terahertz}
\newacronym{mimo}{MIMO}{multiple-input multiple-output}
\newacronym{um-mimo}{UM-MIMO}{ultra-massive multiple-input multiple-output}
\newacronym{pdf}{PDF}{probability density function}
\newacronym{ber}{BER}{bit-error rate}
\newacronym{mg}{MG}{mixture of gamma}
\newacronym{gm}{GM}{Gaussian mixture}
\newacronym{pep}{PEP}{pairwise error probability}
\newacronym{los}{LoS}{line-of-sight}
\newacronym{nlos}{NLoS}{non-LoS}
\newacronym{aosas}{AoSAs}{array of subarrays}
\newacronym{ofdm}{OFDM}{orthogonal frequency division multiplexing}
\newacronym{ml}{ML}{maximum-likelihood}
\newacronym{nf}{NF}{near-field}
\newacronym{ff}{FF}{far-field}
\newacronym{fpga}{FPGA}{field programmable gate arrays}
\newacronym{mmse}{MMSE}{minimum-mean-square-error}
\newacronym{wmmse}{WMMSE}{weighted-MMSE}
\newacronym{svm}{SVM}{support-vector-machine}
\newacronym{sic}{SIC}{successive-interference-cancellation}
\newacronym{qrd}{QRD}{QR decompostion}
\newacronym{wrd}{WRD}{WR decompostion}
\newacronym{sqrd}{SQRD}{sorted-QR decompostion}
\newacronym{air}{AIR}{achievable information rate}
\newacronym{zf}{ZF}{zero-forcing}
\newacronym{nmse}{NMSE}{normalized mean-square error}
\newacronym{upa}{UPA}{uniform planar array}

\newacronym{sd}{SD}{sphere-decoding}
\newacronym{ttd}{TTD}{true-time-delay}
\newacronym{lord}{LORD}{layered orthogonal lattice detector}
\newacronym{g-lord}{G-LORD}{global layered orthogonal lattice detector}
\newacronym{v-lord}{V-LORD}{vector-based LORD}
\newacronym{ssd}{SSD}{subspace detector}
\newacronym{pml}{PML}{punctured-ML}
\newacronym{flops}{FLOPs}{floating-point operations}
\newacronym{lr}{LR}{lattice reduction}
\newacronym{snr}{SNR}{signal-to-noise ratio}
\newacronym{llr}{LLR}{log likelihood ratio}
\newacronym{qpsk}{QPSK}{Quadrature Phase Shift Keying}
\newacronym{sa}{SA}{subarray}
\newacronym{sc}{SC}{single carrier}
\newacronym{mc}{MC}{multi carrier}
\newacronym{ae}{AE}{antenna element}
\newacronym{csi}{CSI}{channel state information}
\newacronym{swm}{SWM}{spherical wave model}
\newacronym{pwm}{PWM}{planar wave model}
\newacronym{hspwm}{HSPWM}{hybrid spherical-planar wave model}
\newacronym{so}{SO}{soft-output}
\newacronym{radd}{RADD}{real addition}
\newacronym{rmul}{RMUL}{real multiplication}
\newacronym{kpi}{KPI}{key performance indicators}
\newacronym{v-blast}{V-BLAST}{vertical-bell laboratories layered space-time}
\newacronym{plr}{PLR}{partial lattice reduction}
\newacronym{sqld}{SQLD}{sorted-QR layered detector}

\newacronym{}{}{}

\newacronym{awgn}{AWGN}{additive white gaussian noise}
\newacronym{cbm}{CBM}{Carrier Based Modulation}
\newacronym{i/o}{I/O}{input and output}
\newacronym{cdf}{CDF}{Cumulative Distribution Function}
\newacronym{cd}{CD}{chase detector}
\newacronym{iid}{i.i.d}{independent and identically distributed}

\newacronym{mmwave}{mmWave}{millimetre waves}
\newacronym{m-psk}{$M$-PSK}{$M$-ary Phase Shift Keying}
\newacronym{m-qam}{$M$-QAM}{$M$-ary Quadrature Amplitude Modulation}
\newacronym{pbm}{PBM}{Pulse Based Modulation}
\newacronym{psd}{PSD}{Power Spectral Density}
\newacronym{cs}{CS}{compressed-sensing}

\newacronym{fd}{FD}{fully-digital}
\newacronym{fc}{FC}{fully-connected}
\newacronym{pc}{PC}{partially-connected}
\newacronym{svd}{SVD}{singular value decomposition}
\newacronym{omp}{OMP}{orthogonal-matching pursuit}
\newacronym{somp}{SOMP}{Simultaneous-OMP}
\newacronym{omp-dr}{OMP-DR}{dictionary reduction OMP}

\newacronym{rf}{RF}{radio frequency}
\newacronym{mpc}{MPC}{multi path component}
\newacronym{fsd}{FSD}{fixed sphere decoding}
\newacronym{lll}{LLL}{
Lenstra–Lenstra–Lovász}
\newacronym{clll}{CLLL}{Complex Lenstra–Lenstra–Lovász}
\newacronym{qam}{QAM}{quadratic amplitude modulation}
\newacronym{aoa}{AoA}{angle of arrival}
\newacronym{aod}{AoD}{angle of departure}
\newacronym{toa}{ToA}{time of arrival}
\newacronym{hiho}{HIHO}{hard input hard output}
\newacronym{siso}{SISO}{soft input soft output}
\newacronym{ldpc}{LDPC}{low-density parity-check}
\newacronym{lte}{LTE}{long term evolution}

\newacronym{pn}{PN}{phase noise}

\usepackage{amsmath} 
\usepackage{amssymb}
\usepackage{amsxtra}
\usepackage{enumitem}
\usepackage{float}
\usepackage{booktabs}
\usepackage[T1]{fontenc}
\usepackage{setspace}
\usepackage{makecell} 

\usepackage[cmintegrals]{newtxmath}
\usepackage{graphicx}
\graphicspath{{Figures/}}
\usepackage{url}
\usepackage{cite}
\usepackage{algpseudocode}

\makeatletter
\def\BState{\State\hskip-\ALG@thistlm}
\makeatother

\usepackage{epstopdf}
\usepackage{amsxtra}
\usepackage{amstext}
\usepackage{array}

\usepackage{amssymb}
\usepackage{latexsym}
\usepackage{dsfont} 
\usepackage{color}
\usepackage{multirow}
\usepackage{float}
\usepackage{hyperref}
\usepackage{tabularx}

\usepackage{tabularx,colortbl}
\usepackage[table]{xcolor}

\usepackage{lscape}
\usepackage{diagbox}

\usepackage{algorithm}
\usepackage{algpseudocode}

\usepackage{mwe}
\usepackage{units}
\usepackage{cases}
\usepackage[ subrefformat=parens,labelformat=parens, caption=false, font=footnotesize]{subfig}
\usepackage{mathtools}

\newcommand{\Hpow}{{\sf H}}
\newcommand{\Tpow}{{\sf T}}
\newcommand{\Invpow}{{\sf -1}}
\newcommand{\Strpow}{{\sf *}}

\newcommand{\mbf}[1]{\mathbf{#1}}
\newcommand{\blue}[1]{{\color{blue}{#1}}} 

\newcommand{\nth}[1]{{#1}{\text{th}}}

\newcommand{\abs}[1]{\left|{#1}\right|}
\newcommand{\norm}[1]{\left\|{#1}\right\|}

\newtheorem{theorem}{Theorem}
\newenvironment{proof}{\noindent\textbf{Proof:} }{\hfill$\blacksquare$\par}

\usepackage{mathtools}

\DeclareFontFamily{U}{mathx}{\hyphenchar\font45}
\DeclareFontShape{U}{mathx}{m}{n}{
      <5> <6> <7> <8> <9> <10>
      <10.95> <12> <14.4> <17.28> <20.74> <24.88>
      mathx10
      }{}
\DeclareSymbolFont{mathx}{U}{mathx}{m}{n}
\DeclareFontSubstitution{U}{mathx}{m}{n}
\DeclareMathAccent{\widecheck}{0}{mathx}{"71}

\newcommand{\LoS}{\mathrm{LoS}}
\newcommand{\NLoS}{\mathrm{NLoS}}

\newcommand{\clu}{\mathrm{clu}}
\newcommand{\ray}{\mathrm{ray}}

\mathchardef\mhyphen="2D
\begin{document}

\title{Performance and Complexity Analysis of Terahertz-Band MIMO Detection}

\author{Hakim~Jemaa,~\IEEEmembership{Graduate Student Member,~IEEE,}
        Simon Tarboush,
        Hadi~Sarieddeen,~\IEEEmembership{Senior Member,~IEEE,}
        Mohamed-Slim~Alouini,~\IEEEmembership{Fellow,~IEEE,}
        and~Tareq~Y.~Al-Naffouri,~\IEEEmembership{Fellow,~IEEE}
\thanks{This work was supported by the Office of Sponsored Research at King Abdullah University of Science and Technology (KAUST) (Award ORA-CRG2021-4695), and the University Research Board at the American University of Beirut (AUB) (Award 104522). Preliminary results were presented at the IEEE Asilomar Conference on Signals, Systems, and Computers \cite{JemaaDetection2022}.  H.~Jemaa, M.-S.~Alouini, and T.~Y.~Al-Naffouri are with the Department of Computer, Electrical and Mathematical Sciences and Engineering, KAUST, Kingdom of Saudi Arabia, 23955-6900 (email: \{hakim.jemaa;  slim.alouini; tareq.alnaffouri\}@kaust.edu.sa). S. Tarboush is with the Faculty of Electrical Engineering and Computer Science, Technical University of Berlin, Germany (email: simon.w.tarboush@gmail.com). H. Sarieddeen is with the Electrical and Computer Engineering Department, AUB, Lebanon (email: hadi.sarieddeen@aub.edu.lb).}
}

\maketitle

\begin{abstract}
Achieving terabit-per-second (Tbps) data rates in terahertz (THz)-band communications requires bridging the complexity gap in baseband transceiver design. This work addresses the signal processing challenges associated with data detection in THz-band \gls{mimo} systems. We begin by analyzing the trade-offs between performance and complexity across various detection schemes and THz channel models, demonstrating significant complexity reduction by leveraging spatial parallelism across subspaces of correlated, typically ill-conditioned THz \gls{mimo} channels. We also derive accurate theoretical bounds on the detection error probability by incorporating THz-specific channel distributions and accounting for mismatches introduced by subspace decomposition. In addition, we propose a variation of subspace detectors that combines channel-matrix sorting, QR decomposition, and puncturing. Furthermore, under wideband THz UM-MIMO systems, we introduce a channel-matrix reuse strategy that minimizes exhaustive computations while maintaining reliable detection performance within a coherence bandwidth. Simulations over accurate THz channels show that the proposed efficient spatial parallelization schemes yield multi-dB performance gains, while the proposed reuse strategy offers significant computational savings with minimal performance degradation.
\end{abstract}


\maketitle

\section{Introduction}
\IEEEPARstart{T}he \gls{thz} band, spanning $\unit[0.1-10]{\gls{thz}}$, is expected to enable future wireless communications~\cite{akyildiz2022terahertz}. THz communications promise \gls{tbps} data rates and novel joint communication-sensing applications~\cite{sarieddeen2020overview}. However, \gls{thz}-band device limitations and high propagation losses limit spectral and energy efficiency~\cite{Jornet2024Evolution}. These constraints can be mitigated by infrastructure enablers such as \gls{um-mimo} antenna arrays, which provide the beamforming gains for range extension~\cite{rajatheva2020white}. Nevertheless, expanding the operating bandwidths and antenna dimensions imposes stringent \gls{thz}-band baseband processing demands, where achieving \gls{tbps} rates requires parallelizable transceiver designs that overcome hardware limitations~\cite{sarieddeen2023bridging2}.

Accurate \gls{thz} channel modeling is crucial for efficient signal processing. \gls{thz} signals undergo significant path loss due to spreading and molecular absorption losses~\cite{tarboush9591285}, which increases with distance and leads to distance-dependent frequency selectivity, even for a \gls{los} scenario. Nonetheless, absorption-free spectral windows provide access to vast bandwidths. Communication via \gls{nlos} paths is possible~\cite{sheikh2022thz}, but even in indoor \gls{thz} environments, these paths are limited, especially with high-gain antennas or massive beamforming~\cite{tarboush9591285}. Recent measurement campaigns indicate that the \gls{gm} distribution effectively models small-scale fading in outdoor \gls{thz} links, capturing multipath-induced peaks in highly correlated channels~\cite{Jemaa2024Performance, papasotiriou2023outdoor}. Additionally, the $\alpha\mhyphen\mu$ distribution accurately characterizes small-scale fading in indoor \gls{thz} environments due to its adaptability to diverse propagation conditions~\cite{papasotiriou2021experimentally}.

\gls{thz}-specific propagation effects intensify with large spatial and spectral dimensions. In the spatial domain, using \gls{um-mimo} can help mitigate hardware and power constraints while balancing complexity and spectral efficiency~\cite{tarboush9591285}. For massive arrays and extremely short wavelengths, far-field models become inaccurate, making the \gls{swm} essential for near-field \gls{thz} communications~\cite{Lu10496996}. In the \gls{swm}, each antenna element experiences a distinct channel response due to variations in both amplitude and phase, unlike the \gls{pwm} in \gls{ff}. Accounting for spatial correlation between array elements, particularly in dense array configurations~\cite{9763525}, complicates MIMO detection. In multicarrier wideband \gls{thz} systems, efficient waveform design is critical, as conventional waveforms like \gls{ofdm} face limitations~\cite{tarboush2022single}. Additionally, wideband \gls{thz} systems suffer from the beam-split effect, where frequency-dependent beam divergence complicates beamforming and beam alignment~\cite{Wang2022HybridBeamforming}. To address these challenges, adaptive multi-wideband waveform designs~\cite{Han2016Multi} and \gls{ttd} elements~\cite{Wang2022HybridBeamforming} have been proposed. Such \gls{thz}-specific characteristics necessitate advanced transceiver designs~\cite{tarboush2022single, Jornet2024Evolution}.

\subsection{Relevant works on MIMO data detection}

\gls{thz}-band system and channel constraints significantly exacerbate the challenge of low-complexity \gls{mimo} data detection~\cite{7244171}. While \gls{ml} detection offers optimal performance~\cite{7472341}, its prohibitive complexity makes it impractical for \gls{um-mimo} systems. Various linear and nonlinear detection algorithms have been developed to balance complexity and performance~\cite{7244171}. Linear detection, though near-optimal in conventional asymmetric massive \gls{mimo} due to channel hardening~\cite{Sarieddeen8765243}, suffers severe performance degradation in highly correlated symmetric \gls{thz} \gls{mimo} channels~\cite{Sarieddeen8765243}. Tree-search-based detectors offer flexible complexity-performance trade-offs~\cite{7244171} but become impractical as system dimensions grow. Approximate message passing (AMP) detectors achieve low complexity and strong performance in large-scale \gls{mimo}~\cite{jeon2015optimality} but struggle under high channel correlation, a key challenge in \gls{thz}-band links~\cite{9763525}, especially in \gls{um-mimo} near-field scenarios.

In recent notable works, a high-throughput data detection algorithm for massive \gls{mimo} \gls{ofdm} systems, which uses coordinate descent for approximate \gls{mmse} within a parallelizable architecture, was proposed in \cite{7755889}. The work in \cite{9148630} introduces a \gls{svm}-based method for channel estimation and data detection in one-bit massive \gls{mimo} systems, effectively mitigating quantization noise. Similarly, \cite{kobayashi2016} proposes an ordered \gls{mmse}-\gls{sic} detection method based on \gls{sqrd}. Furthermore, the \gls{air}-based detector in \cite{hu2017} reduces complexity and latency in \gls{mimo} detection, outperforming traditional methods like K-Best sphere decoding and \gls{zf}. While these studies enhance \gls{mimo} detection in power efficiency, throughput, and performance, further research is required to improve parallelizability and adaptability to diverse channel conditions.

Detection in correlated THz wideband multicarrier systems has been relatively underexplored. Early work \cite{ammari2015analysis} concentrated on Rayleigh fading channel models, which are appropriate for lower-frequency bands but do not fully capture the complexity of wideband \gls{um-mimo} scenarios. More recent studies \cite{9896734} investigated wideband \gls{thz} \gls{mimo} systems, enhancing channel estimation through iterative refinement to mitigate the beam split effect. In particular, \cite{9896734} introduced a 3-D hybrid beamforming framework that integrates \gls{ttd} elements with analog phase shifters, where delay-phase precoding enables frequency-dependent beamforming vectors to improve detection accuracy and suppress interference. Nevertheless, the dependence on \gls{ttd} components substantially increases hardware complexity and power consumption, limiting the feasibility of large-scale deployment in \gls{um-mimo} systems compared to other hybrid beamforming alternatives. A comprehensive comparative study of different detection schemes in wideband \gls{thz} \gls{mimo} systems remains largely unexplored, highlighting a key research gap that this work aims to address.

Achieving \gls{tbps} data rates requires parallelizable transceiver architectures to overcome hardware limitations in integrated circuit clock frequencies, bridging the ``\gls{tbps} gap'' \cite{sarieddeen2023bridging2}. Detectors such as \gls{lord} and \gls{ssd}, which transform channel matrices into parallelizable structures, show promise \cite{Sarieddeen8186206}. To the best of the authors’ knowledge, a comprehensive study on \gls{thz} \gls{um-mimo} wideband data detection remains unexplored, with no prior research addressing this topic in sufficient depth.




\subsection{Contributions}

This work investigates data detection in \gls{thz}-band \gls{mimo} and \gls{um-mimo} systems, considering various \gls{thz} scenarios of distinct challenges and opportunities. We propose solutions addressing both \gls{thz} channel characteristics and baseband processing constraints. The main contributions are: 
\begin{itemize}
    \item We investigate a range of candidate data detection algorithms under diverse \gls{thz} channel conditions, highlighting performance–complexity trade-offs and the importance of parallelizability in reducing complexity overhead. In addition, we examine the impact of channel estimation accuracy and hybrid precoding architectures by applying the state-of-the-art schemes for both sparse channel estimation and hybrid-precoding on detector performance. The analysis of \gls{mimo} detection is further extended to account for beam-split effects in wideband scenarios, where we quantify the associated performance degradation.
    \item We advocate for subspace detection algorithms as a means to improve both parallelizability and performance. In particular, under ill-conditioned, \gls{los}-dominant \gls{thz} channels, \gls{ssd} achieves substantial performance gains of approximately \unit[10]{dB} over \gls{lord}, while attaining performance comparable to the suboptimal subspace detector, \gls{v-lord}, in ill-conditioned \gls{thz} \gls{um-mimo} \gls{los} channels.
    \item We derive a closed-form theoretical lower bound on the \gls{pep} for \gls{thz}-band \gls{ml} \gls{mimo} detection by approximating the Frobenius norm squared distribution under the $\alpha\mhyphen\mu$ \gls{thz} fading model. This bound is shown to remain valid under correlated \gls{mimo} channels such as the Kronecker model.
    \item We propose the \gls{sqld}, which integrates channel-matrix sorting and puncturing to mitigate ill-conditioned \gls{thz} channel effects. By combining the strengths of \gls{lord}, \gls{ssd}, and sorting, \gls{sqld} enhances detection accuracy and promotes parallelizability.
    \item We propose a channel-matrix reuse strategy for wideband \gls{thz} systems that reduces \gls{flops}, processing time, and power consumption by reusing \gls{qrd} computations obtained from a subset of subcarriers across the remaining subcarriers.
    \item We investigate the performance degradation of subspace detectors caused by channel puncturing and show, using empirical bounds, that this loss can be alleviated in ill-conditioned \gls{thz} channels.
\end{itemize}

\subsection{Organization and notation}

The paper is structured as follows: Sec.~\ref{sec:sysmodel} introduces the system and channel models. Sec.~\ref{sec:cand_data_det} formulates the data detection problem, reviews candidate linear and nonlinear detectors, and presents the proposed \gls{sqld} algorithm. Sec.~\ref{sec:MIMO_perf_analysis} derives the error probability bounds for \gls{ml} detection under \gls{thz} conditions. Sec.~\ref{sec:complexity} examines detection complexity and introduces a low-complexity \gls{thz} wideband design with a channel-matrix reuse strategy. Sec.~\ref{sec:sim_res_disc} presents simulation results and discussions. 

We use the following notation throughout the paper. Non-bold lower and upper case letters ($a, A$) denote scalars, bold lower case letters ($\mbf{a}$) denote vectors, and bold upper case letters ($\mbf{A}$) denote matrices. The superscripts ${(\cdot)}^\Tpow$, ${(\cdot)}^\Strpow$, ${(\cdot)}^\Hpow$, and ${(\cdot)}^\Invpow$ represent the transpose, conjugate, hermitian, and inverse operators, respectively. The $\abs{\cdot}$ operator can represent the absolute value of $a$ and the absolute value of each entry of $\mbf{a}$; $\norm{\mbf{a}}$ is the Euclidean norm of $\mbf{a}$. $\mbf{0}_{N,1}$, $\mbf{0}_{N}$, and $\mbf{I}_N$ are the zero vector of size $N\times 1$, zero matrix of size $N\times N$, and identity matrix of size $N\times N$, respectively. $\mathrm{Tr}(\cdot)$ represents the trace operator. $j=\sqrt{-1}$ denotes the imaginary unit. $\mathcal{CN}(\mbf{a},\mbf{A})$ denotes a complex circularly symmetric Gaussian random vector of mean $\mbf{a}$ and covariance $\mbf{A}$. $\mathbb{E}[\cdot]$ is the expectation operator and $\mathrm{Pr}(\cdot)$ is the probability operator. $\ln(\cdot)$ represents the natural logarithm, $\Gamma(.)$ is the Gamma function, and $\Gamma(.,.)$ is the incomplete gamma function defined as $\Gamma(\xi, u)=\int_0^u t^{\xi-1} e^{-t} d t$. 
$Q(x)=\frac{1}{\sqrt{2 \pi}} \int_x^{\infty} e^{-\frac{u^2}{2}} d u$ is the $Q\mhyphen$function and ${Q}(x) \!=\! \frac{1}{2} \mathrm{erfc}(\frac{x}{\sqrt{2}})$ where $\text{erfc}$ is the complementary error function. $G_{p, q}^{m, n}\left[z \left\lvert \begin{smallmatrix}a_1, \ldots, a_p \\ b_1, \ldots, b_q\end{smallmatrix} \right.\right]$ represents the Meijer G-function \cite[eq. (9.301)]{prudnikov1986integrals} while $H_{p, q}^{m, n}\left[z \left\lvert \begin{smallmatrix}(a_1, b_1), \ldots, (a_p, b_p) \\ (c_1, d_1), \ldots, (c_p, d_p)\end{smallmatrix} \right.\right]$ represents the Fox H-function \cite[eq. (1.1.1)]{kilbas2004h}. 

\maketitle
\section{System and channel models}
\label{sec:sysmodel}

\subsection{System and received-signal model}
\begin{figure*}[ht]
  \centering
  \includegraphics[width =0.93\linewidth]{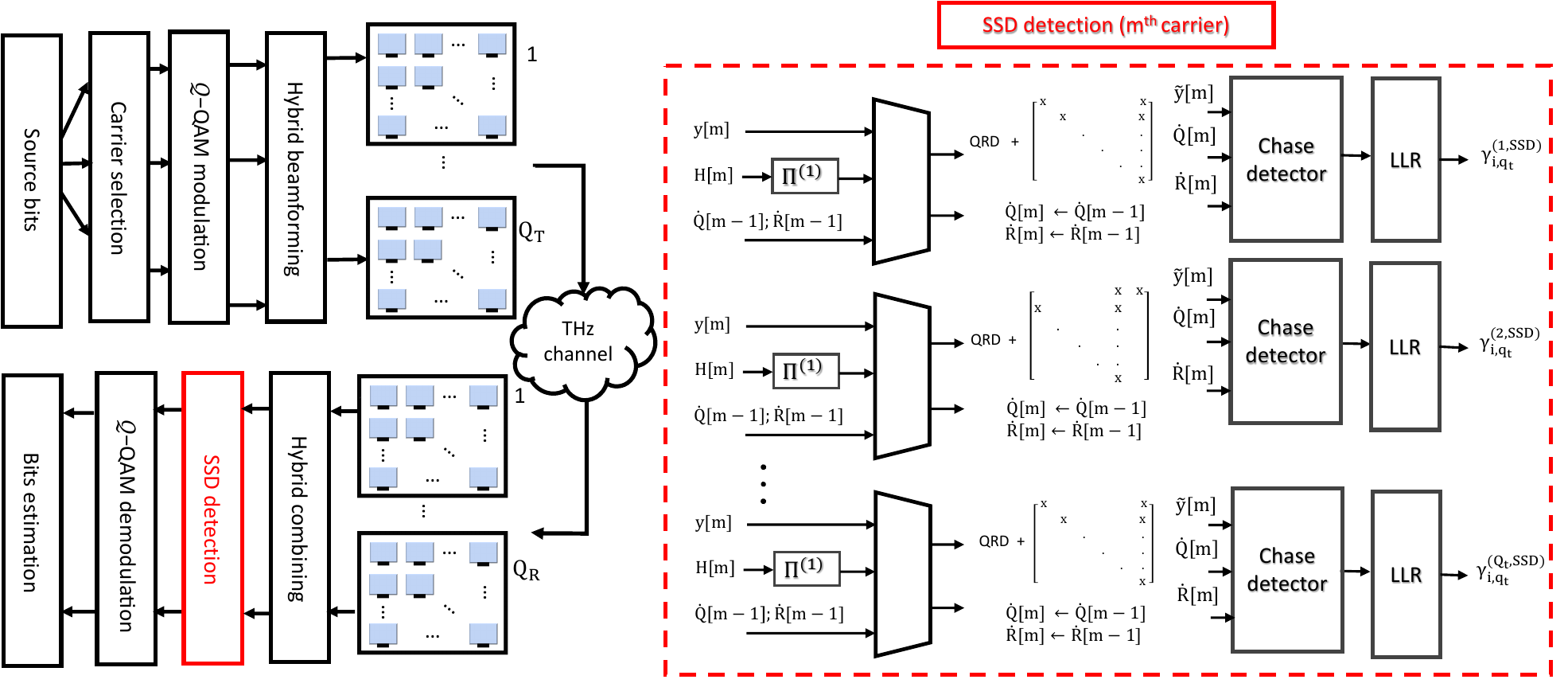}
  \caption{Block diagram of a wideband \gls{um-mimo} \gls{thz} communication system adopting an \gls{aosas} architecture and subspace detection.}
  \label{fig:aosa_tx_rx}
\end{figure*}

We adopt a multicarrier \gls{um-mimo} system for \gls{thz} communications~\cite{tarboush2022single}, leveraging a hybrid-beamforming \gls{aosas} architecture for spatial multiplexing and beamforming gains~\cite{tarboush9591285}. The transmitter and receiver consist of $Q_{\mathrm t}=M_{\mathrm t}N_{\mathrm t}$ and $Q_{\mathrm r}=M_{\mathrm r}N_{\mathrm r}$ \glspl{sa}, respectively, arranged on $M_{\mathrm t}N_{\mathrm t}$ and $M_{\mathrm r}N_{\mathrm r}$ grids. Each \gls{sa} is a \gls{upa} of $Q^2$ \gls{ae}s with intra-\gls{sa} spacings $\delta_m,\delta_n$; we set $\delta_m=\delta_n\triangleq\delta=\lambda/2$.\footnote{We set $\delta=\lambda/2$ to mitigate the mutual coupling (MC) effect among closely spaced \gls{ae}s; an in-depth MC-aware analysis is left to future work.} The center-to-center spacings between adjacent \gls{sa}s are $\Delta_{\mathrm{t},m},\Delta_{\mathrm{t},n}$ at the transmitter and $\Delta_{\mathrm{r},m},\Delta_{\mathrm{r},n}$ at the receiver, defined compactly by
$\Delta_{\mathrm{t},m}=k_{\mathrm{t},m}Q_m\delta,\ \Delta_{\mathrm{t},n}=k_{\mathrm{t},n}Q_n\delta$ and
$\Delta_{\mathrm{r},m}=k_{\mathrm{r},m}Q_m\delta,\ \Delta_{\mathrm{r},n}=k_{\mathrm{r},n}Q_n\delta$,
with integers $k_{\cdot}\!\ge\!1$. 

Let $M$ denote the number of subcarriers, indexed by $m\in\{0,\ldots,M-1\}$. The received signal at the $m$-th subcarrier, $\mathbf{y}[m] \in \mathbb{C}^{N_\mathrm{s} \times 1}$, is \cite{tarboush2022single}
\begin{equation}
\label{eq:sys_model_firstline}
\mathbf{y}[m]=\mathbf{W}_{\mathrm{BB}}^\Hpow[m]\mathbf{W}_{\mathrm{RF}}^\Hpow\!\Big(\widecheck{\mathbf{H}}[m]\mathbf{F}_{\mathrm{RF}}\mathbf{F}_{\mathrm{BB}}[m]\mathbf{x}[m]+\widecheck{\mathbf{n}}[m]\Big),
\end{equation}
where $\mathbf{x}[m]\in\mathbb{C}^{N_{\mathrm s}\times 1}$ is the information-bearing vector with $N_{\mathrm s}\le \min(Q_{\mathrm t},Q_{\mathrm r})$ (entries are drawn from a \gls{qam} constellation $\mathcal{X}$); $\widecheck{\mathbf{H}}[m]\in\mathbb{C}^{Q_{\mathrm r}Q^2\times Q_{\mathrm t}Q^2}$ is the \gls{ae}-domain \gls{um-mimo} channel matrix; $\mathbf{F}_{\mathrm{RF}}\in\mathbb{C}^{Q_{\mathrm t}Q^2\times Q_{\mathrm t}}$ and $\mathbf{W}_{\mathrm{RF}}\in\mathbb{C}^{Q_{\mathrm r}Q^2\times Q_{\mathrm r}}$ are the frequency-flat analog \gls{rf} precoder/combiner; $\mathbf{F}_{\mathrm{BB}}[m]\in\mathbb{C}^{Q_{\mathrm t}\times N_{\mathrm s}}$ and $\mathbf{W}_{\mathrm{BB}}[m]\in\mathbb{C}^{Q_{\mathrm r}\times N_{\mathrm s}}$ are the per-subcarrier digital baseband precoder/combiner; $\widecheck{\mathbf{n}}[m]\sim\mathcal{CN}\!\big(\mathbf{0}_{Q_{\mathrm r}Q^2},\sigma^2\mathbf{I}_{Q_{\mathrm r}Q^2}\big)$ is the pre-combining noise; and $\mathbf{n}[m]=\mathbf{W}_{\mathrm{BB}}^\Hpow[m]\mathbf{W}_{\mathrm{RF}}^\Hpow\widecheck{\mathbf{n}}[m]\in\mathbb{C}^{N_{\mathrm s}\times 1}$ is the post-combining noise. Without loss of generality, we set $N_{\mathrm{s}} = Q_{\mathrm{t}} = Q_{\mathrm{r}}$ for the remainder of this work.    

We adopt a sub-connected hybrid architecture with one \gls{rf} chain per \gls{sa}. Consequently, the analog \gls{rf} precoder/combiner are block diagonal with constant-modulus entries:
\begin{align}
\label{eq:block_RF}
\mathbf{F}_{\mathrm{RF}}&=\operatorname{blkdiag}\{\mathbf{f}_1,\ldots,\mathbf{f}_{Q_{\mathrm t}}\},\quad
\mathbf{W}_{\mathrm{RF}}=\operatorname{blkdiag}\{\mathbf{w}_1,\ldots,\mathbf{w}_{Q_{\mathrm r}}\},\\[-0.5mm]
&\mathbf{f}_q,\mathbf{w}_p\in\mathbb{C}^{Q^2\times 1},\;\;q=1,\ldots,Q_{\mathrm t},\;\;p=1,\ldots,Q_{\mathrm r},\nonumber\\[-0.5mm]
&|(\mathbf{f}_q)_i|=|(\mathbf{w}_p)_j|=1/\sqrt{Q^2},\;\;i,j=1,\ldots,Q^2.\nonumber
\end{align}
(The receiver \gls{rf}/baseband combiners obey the same structure, assumptions, and constraints as the transmitter.)
We partition $\widecheck{\mathbf{H}}[m]$ into $Q_{\mathrm r}\times Q_{\mathrm t}$ \gls{sa} blocks, each of size $Q^2\times Q^2$:
\begin{align}
\widecheck{\mathbf{H}}[m]=
\begin{bmatrix}
\widecheck{\mathbf{H}}_{1,1}[m] & \cdots & \widecheck{\mathbf{H}}_{1,Q_{\mathrm t}}[m]\\
\vdots & \ddots & \vdots\\
\widecheck{\mathbf{H}}_{Q_{\mathrm r},1}[m] & \cdots & \widecheck{\mathbf{H}}_{Q_{\mathrm r},Q_{\mathrm t}}[m]
\end{bmatrix}\!,\;\;\widecheck{\mathbf{H}}_{p,q}[m]\!\in\!\mathbb{C}^{Q^2\times Q^2}.
\end{align}
After \gls{rf} precoding/combining, the post-\gls{rf} \gls{sa}-level effective channel is
\begin{align}
\label{eq:H_SA_def}
\widehat{\mathbf{H}}_{\mathrm{SA}}[m]&\triangleq \mathbf{W}_{\mathrm{RF}}^\Hpow\,\widecheck{\mathbf{H}}[m]\,\mathbf{F}_{\mathrm{RF}}
=\big[\,\hat h_{p,q}[m]\,\big]_{p=1,\ldots,Q_{\mathrm r}}^{q=1,\ldots,Q_{\mathrm t}}\in\mathbb{C}^{Q_{\mathrm r}\times Q_{\mathrm t}},\\
\hat h_{p,q}[m]&=\mathbf{w}_p^\Hpow\,\widecheck{\mathbf{H}}_{p,q}[m]\,\mathbf{f}_q\in\mathbb{C}.
\end{align}
After digital precoding/combining, the detection (baseband) channel is
\begin{align}
\label{eq:H_detect}
\mathbf{H}[m]\triangleq \mathbf{W}_{\mathrm{BB}}^\Hpow[m]\,\widehat{\mathbf{H}}_{\mathrm{SA}}[m]\,\mathbf{F}_{\mathrm{BB}}[m]\in\mathbb{C}^{N_{\mathrm s}\times N_{\mathrm s}}.
\end{align}
Accordingly, \eqref{eq:sys_model_firstline} becomes
\begin{align}
\label{eq:sys_model_secondline}
\mathbf{y}[m]=\mathbf{H}[m]\mathbf{x}[m]+\mathbf{n}[m],\quad
\mathbf{n}[m]=\mathbf{W}_{\mathrm{BB}}^\Hpow[m]\mathbf{W}_{\mathrm{RF}}^\Hpow\,\widecheck{\mathbf{n}}[m],
\end{align}
which is the input–output relation used for detection.

\subsection{THz MIMO channel modeling}
\label{subsec:thz_channel_models}
For most of the analysis, we assume perfect \gls{csi} at the receiver and generate channels using the TeraMIMO simulator~\cite{tarboush9591285}, which captures \gls{los}/\gls{nlos} clusters, molecular absorption, beam split, \gls{nf} and \gls{ff} propagation, and \gls{aosas} geometry. We later evaluate in depth the impact of imperfect \gls{csi} in the simulation section. The equivalent frequency-domain baseband channel between the $q_{\mathrm t}$-th Tx \gls{sa} and the $q_{\mathrm r}$-th Rx \gls{sa} at subcarrier $m$ is \cite{tarboush9591285}
\begin{align}
\label{eq:ch_sa_freq_domain}
& h_{q_{\mathrm r},q_{\mathrm t}}[m]
=\alpha^{\LoS}(m)\,\underbrace{\mathbf{w}_{q_{\mathrm r}}^{\Hpow}\mathbf{a}_{\mathrm r}(\mbf{\Phi}_{\mathrm r})}_{\triangleq\,G_{\mathrm r}(\mbf{\Phi}_{\mathrm r})}\,
\underbrace{\mathbf{a}_{\mathrm t}^{\Hpow}(\mbf{\Phi}_{\mathrm t})\mathbf{f}_{q_{\mathrm t}}}_{\triangleq\,G_{\mathrm t}(\mbf{\Phi}_{\mathrm t})}\,
e^{-j 2\pi \frac{m B}{M} \tau^{\LoS}}\\
&\quad+\sum_{c=1}^{N_{\clu}}\sum_{\ell=1}^{N_{\ray}^{c}}
\alpha_{c,\ell}^{\NLoS}(m)\,G_{\mathrm r}(\mbf{\Phi}_{\mathrm r,c,\ell})\,G_{\mathrm t}(\mbf{\Phi}_{\mathrm t,c,\ell})\,
e^{-j 2\pi \frac{m B}{M} \tau^{\NLoS}_{c,\ell}},
\end{align}
where $\mathbf{a}_{\mathrm t}(\cdot),\mathbf{a}_{\mathrm r}(\cdot)\in\mathbb{C}^{Q^2}$ are the \gls{sa} \gls{upa} steering vectors, $\mathbf{f}_{q_{\mathrm t}},\mathbf{w}_{q_{\mathrm r}}\in\mathbb{C}^{Q^2}$ are the constant-modulus \gls{rf} beamformers of the $q_{\mathrm t}$-th Tx \gls{sa} and $q_{\mathrm r}$-th Rx \gls{sa} (cf. \eqref{eq:block_RF}), and $h_{q_{\mathrm r},q_{\mathrm t}}[m]$ is exactly the $(q_{\mathrm r},q_{\mathrm t})$ entry of $\widehat{\mathbf{H}}_{\mathrm{SA}}[m]$ in \eqref{eq:H_SA_def}. Here $B$ is the total bandwidth, $N_{\clu}$ is the number of clusters, $N_{\ray}^{c}$ the number of rays in cluster $c$, $\alpha^{\LoS},\alpha_{c,\ell}^{\NLoS}$ are \gls{los}/\gls{nlos} path gains (including spreading and molecular absorption), $\tau^{\LoS},\tau^{\NLoS}_{c,\ell}$ are the corresponding delays, and $\mbf{\Phi}$ collects the \gls{aod} and \gls{aoa}. This \gls{sa}-level representation encompasses indoor \gls{nlos}, outdoor \gls{los}-dominant, \gls{nlos}-assisted, and \gls{los}-only scenarios \cite{tarboush9591285,tarboush2022single}.

We consider four scenarios representative of diverse \gls{thz} environments and map them to two modeling tracks. The first track employs a TeraMIMO-based stochastic channel for indoor clustered and \gls{los}-only links \cite{tarboush9591285}, while the second uses a \gls{pdf}-based small-scale fading model for indoor and outdoor \gls{nlos} conditions \cite{papasotiriou2021experimentally,papasotiriou2023outdoor}. For the first scenario (indoor clustered multipath), we generate channels using TeraMIMO \cite{tarboush9591285}, which captures \gls{thz}-specific effects such as molecular absorption, beam split, 3-D geometry, and \gls{aosas} architecture. The per-\gls{sa} frequency-domain channel between the $q_{\mathrm t}$-th Tx \gls{sa} and the $q_{\mathrm r}$-th Rx \gls{sa} is the scalar $h_{q_{\mathrm r},q_{\mathrm t}}[m]$ in \eqref{eq:ch_sa_freq_domain}; its path-gain terms and angles are drawn according to the TeraMIMO indoor model (clusters/rays with distance– and frequency–dependent statistics). For the second scenario (indoor), we utilize the $\alpha\mhyphen\mu$ distribution to model the magnitude of small-scale fading, as validated through measurements in \cite{papasotiriou2021experimentally}. The \gls{pdf} of an $\alpha\mhyphen\mu$-distributed random variable $X$ is \cite{5090420}
\begin{align}
\label{eq:alphamu_dist}
    f_{X}(x)=\frac{\alpha \beta^{\alpha \mu} x^{\alpha \mu-1}}{\bar{X}^{\alpha \mu} \Gamma(\mu)}e^{-\left(\frac{\beta x}{ \bar{X}}\right)^\alpha },
\end{align}
where $\alpha\!>\!0$ is a fading parameter, $\mu$ represents the normalized variance of the fading channel envelope, $\bar{X}$ indicates the mean value of the fading channel envelope, and $\beta = \frac{\Gamma\left(\mu+\frac{1}{\alpha}\right)}{\Gamma(\mu)}$. The third scenario that of outdoor channels where the magnitude of small-scale fading is modeled using a \gls{mg} distribution, following a \gls{pdf} defined by \cite{papasotiriou2023outdoor}
\begin{align}
\label{MGdist}
    f_{X}(x)=\sum_{i=1}^K w_i \frac{\zeta_i^{\beta_i} x^{\beta_i-1} e^{-\zeta_i x}}{\Gamma\left(\beta_i\right)} = \sum_{i=1}^K \alpha _i x^{\beta _i -1} e^{-\zeta _i x},  x\ge 0,
\end{align}
where $K$ and $w_i$ represent the number of \gls{mg} components and their weights, respectively, and $\alpha_i=w_i\zeta_i^{\beta_i}/\Gamma\left(\beta_i\right)$, with $\zeta _i$ and $\beta _i$ being the scale and shape of the $\nth{i}$ component; the weights satisfy $\sum_{i=1}^K w_i \!=\! 1$. Finally, we use the \gls{los} configuration of TeraMIMO for the fourth scenario (\gls{los}-only) and vary the array geometry/orientation to study the effect of antenna spatial tuning on channel orthogonality at the \gls{sa} level. The choice of TeraMIMO versus \gls{pdf}-based modeling is scenario-driven: the former offers geometry-consistent clustered channels (indoor or \gls{los}) with \gls{aosas} structure and \gls{thz} specifics; the latter provides a compact way to emulate empirically observed \gls{nlos} fading envelopes. Further parameter details are summarized in Sec.~\ref{sec:sim_res_disc}.

\subsection{Wave propagation Models}
The steering vectors in \eqref{eq:ch_sa_freq_domain} depend on the propagation regime determined by the array apertures and the communication distance. A common boundary is the Rayleigh distance
\begin{align}
\label{eq:rayleigh}
d_{\mathrm{Rayleigh}} \;=\; \frac{2\,(D_{\mathrm t}+D_{\mathrm r})^{2}}{\lambda},
\end{align}
where $D_{\mathrm t}$ and $D_{\mathrm r}$ denote the Tx/Rx aperture sizes and $\lambda$ is the wavelength. For communication distances larger than $d_{\mathrm{Rayleigh}}$, the \gls{pwm}, which assumes planar wavefronts, provides an accurate representation \cite{tarboush9591285}. When the distance becomes comparable to or smaller than $d_{\mathrm{Rayleigh}}$, the wavefront curvature and amplitude variation across the aperture become significant, and the \gls{swm} must be employed to capture these \gls{nf} effects. At \gls{thz} frequencies and with the use of \gls{aosas} architectures, the classical Rayleigh criterion may not fully capture the actual boundary between \gls{nf} and \gls{ff}: even for inter-array distances exceeding $d_{\mathrm{Rayleigh}}$, individual \gls{sa}s can still experience spherical wavefronts relative to one another~\cite{tarboush9591285}. Applying the full \gls{swm} across all \gls{ae}s becomes computationally prohibitive. To balance modeling accuracy and complexity, we adopt the \gls{hspwm}, which represents the propagation within each \gls{sa} using planar steering vectors while describing the propagation between different \gls{sa}s using spherical steering vectors. Accordingly, in~\eqref{eq:ch_sa_freq_domain}, the steering vectors $\mathbf{a}_{\mathrm{t}}(\cdot)$ and $\mathbf{a}_{\mathrm{r}}(\cdot)$ are replaced by their hybrid counterparts that account for inter-\gls{sa} curvature while preserving compact planar responses within each \gls{sa}. This formulation naturally aligns with the geometry of large \gls{aosas} architectures without incurring the full complexity of the \gls{swm} at every \gls{ae}. The \gls{hspwm} achieves nearly the same modeling accuracy as the full \gls{swm} with substantially lower computational burden, as verified in~\cite{Tarboush2024Cross}. Hence, the \gls{hspwm} is adopted throughout this work as the propagation model for \gls{aosas}-based \gls{um-mimo} \gls{thz} systems.

\section{Reference and Proposed Data Detectors}
\label{sec:cand_data_det}

We seek \gls{mimo} detectors that separate spatially multiplexed streams for high-throughput, low-latency communication over correlated \gls{thz} channels under stringent \gls{tbps} baseband constraints. In \gls{los}-dominant \gls{thz} scenarios, high channel correlation renders linear detectors ineffective. Adapting nonlinear detectors to search over a subset of candidate vectors in highly parallelizable architectures is preferred \cite{sarieddeen2023bridging2}.  

\subsection{The optimal ML detector}
Under equiprobable priors and assuming Gaussian noise, the optimal \gls{ml} detector solves for
\begin{align}
\label{ML_equation}
\mathbf{\hat{x}}^{\mathrm{ML}}=\underset{\mathbf{x} \in \bar{\mathcal{X}}}{\arg \min d(\mathbf{x})} =\underset{\mathbf{x} \in \bar{\mathcal{X}}}{\arg \min }\|\mathbf{y}-\mathbf{H x}\|_{2}^{2},
\end{align}
where $\bar{X}\!=\!\mathcal{X}^{Q_{\mathrm{t}}}$ is the finite lattice of all candidate transmitted symbol vectors and $d(\mathbf{x}) \triangleq\|\mathbf{y}-\mathbf{H} \mathbf{x}\|_{2}^{2}$ is an Euclidean distance metric. 
Generating \gls{so} detection information in the form of a posteriori bit \gls{llr}s enhance performance when followed by soft channel-code decoding. Let $\mathbf{c} \in \{0,1\}^{\log_2{|\mathcal{X}|} \times Q_{\mathrm{t}}}$ be the bit representation of $\mathbf{x}$; $c_{i, q_{\mathrm{t}}}$ is the $\nth{i}$ bit of the $q_{\mathrm{t}}$th symbol, $q_{\mathrm{t}} \!\in\!\{1,\cdots,Q_{\mathrm{t}}\}$. Then, the \gls{llr}, $\gamma_{i,q_{\mathrm{t}}}$, of $c_{i,q_{\mathrm{t}}}$ is defined as
\begin{align}
\gamma\left(c_{i,q_{\mathrm{t}}}\right|\mathbf{y}) \!\triangleq\! \gamma_{i,q_{\mathrm{t}}} \!\triangleq\! \ln \left({\mathrm{Pr}\left(c_{i,q_{\mathrm{t}}}\!=\!1 | \mathbf{y}, \mathbf{H}\right)}/{\mathrm{Pr}\left(c_{i,q_{\mathrm{t}}}\!=\!0 | \mathbf{y}, \mathbf{H}\right)}\right).
\end{align}
A reduced complexity max-log \gls{llr} approximation is
\begin{align}
\label{eq:maxlog_LLR_ML}
\gamma_{i,q_{\mathrm{t}}}^\mathrm{ML}\!=\!\left\{\begin{array}{ll}d^{\mathrm{ML}}-d_{i,q_{\mathrm{t}}}^{\overline{\mathrm{ML}}} & \text { if } {c}_{i,q_{\mathrm{t}}}^{\mathrm{ML}}=0 \\
    d_{i,q_{\mathrm{t}}}^{\overline{\mathrm{ML}}}-d^{\mathrm{ML}} & \text { if }{c}_{i,q_{\mathrm{t}}}^{\mathrm{ML}}=1,
\end{array}\right.
\end{align}
where the \gls{ml} and counter \gls{ml} distances are, respectively, 
\begin{align}
    d^{\mathrm{ML}}\!=\!\|\mathbf{y}-\mathbf{H} \mathbf{x}^{\mathrm{ML}}\|_{2}^{2}, \ \text{and}\ 
d_{i,q_{\mathrm{t}}}^{\overline{\mathrm{ML}}}\!=\! \min_{\mathbf{x} \in \bar{\mathcal{X}}_{i, q_{\mathrm{t}}}^{\left(c_{i,q_{\mathrm{t}}}^{\overline{\mathrm{ML}}}\right)}}\|\mathbf{y}-\mathbf{H} \mathbf{x}\|_{2}^{2},
\end{align}
where ${\bar{\mathcal{X}}}_{i,q_{\mathrm{t}}}^{(0)}\!=\!\left\{\mbf{x} \!\in\! \bar{\mathcal{X}}\!:{c}_{i,q_{\mathrm{t}}}\!=\!0\right\}$, $\bar{\mathcal{X}}_{i,q_{\mathrm{t}}}^{(1)}\!=\!\left\{\mbf{x} \!\in\! \bar{\mathcal{X}}\!: {c}_{i,q_{\mathrm{t}}}\!=\!1\right\}$, ${c}_{i,q_{\mathrm{t}}}^{\mathrm{ML}}$ is the corresponding bit in $\mathbf{c}^{\mathrm{ML}}$, and $c_{i,q_{\mathrm{t}}}^{\overline{\mathrm{ML}}}$ is its complement.

\subsection{Linear detectors}
A low-complexity linear \gls{zf} detector decouples interfering symbols by multiplying the received signal with the pseudo-inverse of the channel, $\mbf{y}^\mathrm{ZF}=\left(\mbf{H}^{\Hpow} \mbf{H}\right)^{-1}\mbf{H}^{\Hpow} \mbf{y}$. However, large matrix inversions remain computationally demanding. The recovered symbols are quantized to nearest constellation points. The \gls{zf} SOs are
\begin{equation}
\gamma_{i,q_{\mathrm{t}}}^\mathrm{ZF}=\frac{1}{{\sigma_{i}^{\mathrm{ZF}}}^2}\left(\min _{x_{i} \in \mathcal{X}_{i}^{ (1)}}\!\abs{\hat{y}_{i}^{\mathrm{ZF}}-x_{i}}^{2}-\min _{x_{i} \in \mathcal{X}_{i}^{(0)}}\!\abs{\hat{y}_{i}^{\mathrm{ZF}}-x_{i}}^{2}\right),
\end{equation}
where $\mathcal{X}_{i}^{(0)}\!=\!\left\{x \!\in\! {\mathcal{X}}\!:{c}_{i,q_{\mathrm{t}}}\!=\!0\right\}$ and $\mathcal{X}_{i}^{(1)}\!=\!\left\{x \!\in\! {\mathcal{X}}\!: {c}_{i,q_{\mathrm{t}}}\!=\!1\right\}$ are subsets of the \gls{qam} constellation, and ${\sigma_{i}^{\mathrm{ZF}}}^2$ is the scaled post-filtering noise variance. For a Rayleigh-fading channel with \gls{iid} entries, since $\mathbb{E}[\mathbf{n} \mathbf{n}^{\Hpow}] = \sigma^2 \mathbf{I}_{Q_{\mathrm{r}}}$, the post-filtering noise variance for the $i^{\mathrm{th}}$ stream is the $i^{\mathrm{th}}$ diagonal entry of its covariance matrix, leading to ${\sigma_{i}^{\mathrm{ZF}}}^2 \!=\! \sigma^2\left(\mathbf{h}_i^{\Hpow} \mathbf{h}_i\right)^{-1}$. However, the correlation between entries results in colored noise of covariance matrix $\sigma^2\mathbb{E}[(\mathbf{H}^{\Hpow}\mathbf{H})^{-1}]$.

\subsection{Lattice-reduction-aided detectors}
Through \gls{lr}, detection is reformulated as a closest vector search on an infinite lattice with a more orthogonal basis, improving subsequent linear detectors. We employ the \gls{lll} algorithm \cite{4787140} and use Gram-Schmidt orthogonalization to generate an orthogonal basis $\mbf{H}^\mathrm{LR}$ spanning the same space as $\{\mathbf{h}_{1}, \ldots, \mathbf{h}_{Q_{\mathrm{t}}}\}$. Computing the \gls{lr}-aided \gls{zf} \gls{llr}s, denoted $\gamma_{i,q_{\mathrm{t}}}^\mathrm{LR}\!=\!\gamma\left(c_{i,q_{\mathrm{t}}}|\mbf{y}^\mathrm{LR},\mbf{H}^\mathrm{LR}\right)$, involves lattice transformations, imperfect basis orthogonalization, and nearest neighbor quantization, which can degrade performance. Nevertheless, LR-aided detectors are good benchmarks for ill-conditioned THz channels.

\subsection{Tree-search-based detectors}

Balancing complexity and performance in nonlinear detectors requires searching a reduced set of candidate symbol vectors, often via \gls{sd} variants. The search radius within the lattice dictates both complexity and search space. K-Best sphere decoders \cite{1603705} retain the best $K$ paths in a tree-based search (following \gls{qrd}), ensuring fixed complexity and latency. Similar to \eqref{eq:maxlog_LLR_ML}, the K-Best \gls{llr}s are expressed as
\begin{equation}
    \gamma_{i,q_{\mathrm{t}}}^\mathrm{K-Best} = \begin{cases}
        d^{\mathrm{K-Best}} - d_{i,q_{\mathrm{t}}}^{\overline{\mathrm{K-Best}}} & \text{if } c_{i,q_{\mathrm{t}}}^{\mathrm{K-Best}} = 0 \\
        d_{i,q_{\mathrm{t}}}^{\overline{\mathrm{K-Best}}} - d^{\mathrm{K-Best}} & \text{if } c_{i,q_{\mathrm{t}}}^{\mathrm{K-Best}} = 1,
    \end{cases}
\end{equation}
where $d^{{\mathrm{K-Best}}}\!=\! \min_{\mathbf{x} \in \mathcal{S}}\|\tilde{\mathbf{y}}\!-\!\mathbf{R} \mathbf{x}\|_{2}^{2}$, $d_{i,q_{\mathrm{t}}}^{\overline{\mathrm{K-Best}}}\!=\! \min_{\mathbf{x} \in \mathcal{S}_{i, q_{\mathrm{t}}}^{\left(c_{i,q_{\mathrm{t}}}^{\overline{\mathrm{K-Best}}}\right)}}\|\tilde{\mathbf{y}}\!-\!\mathbf{R} \mathbf{x}\|_{2}^{2}$, and $\mathcal{S}\!\subset\!\bar{\mathcal{X}}$ is the set of candidate $\mbf{x}$ vectors populated in a search routine ($\abs{\mathcal{S}}\!=\!K^{\mathrm{Best}}$); $\mathcal{S}_{i,q_{\mathrm{t}}}^{(0)}\!=\!\left\{\mbf{x} \!\in\! \mathcal{S}\!:{c}_{i,q_{\mathrm{t}}}\!=\!0\right\}$ and $\mathcal{S}_{i,q_{\mathrm{t}}}^{(1)}\!=\!\left\{\mbf{x} \!\in\! \mathcal{S}\!: {c}_{i,q_{\mathrm{t}}}\!=\!1\right\}$. 


\subsection{Subspace detectors}
\label{sec:LORD}

Subspace detectors \cite{Sarieddeen8186206} decompose the channel into multiple subspaces, each associated with a limited symbol set. A key example is \gls{lord}, which employs \gls{qrd} for pre-processing and restricts the search to the root layer symbol with projections onto others \cite{Sarieddeen8186206}. This process iterates over $l\!\in\!\{1,2,\dots,Q_{\mathrm{t}}\}$ for different channel permutations, $\Pi^{(l)}$, yielding diverse root-layer symbols. Each decomposition functions as an independent \gls{cd}. Let $\mathbf{Q}^{(l)} \mathbf{R}^{(l)}=\mathbf{H} \Pi^{(l)}$, where $\mathbf{H}\Pi^{(l)}\!=\!\left[\mathbf{h}_{1} \ldots \mathbf{h}_{l-1} \mathbf{h}_{l+1} \ldots \mathbf{h}_{Q_{\mathrm{t}}} \mathbf{h}_{l}\right]$. Similarly permuting $\mbf{x}$ into $\tilde{\mathbf{x}}^{(l)}$, the received vector is expressed as
\begin{equation}
\label{qld_sys}    \tilde{\mathbf{y}}^{(l)}\!=\!\mathbf{Q}^{(l)^{\Hpow}} \mathbf{y}\!=\!\mathbf{R}^{(l)} \Pi^{(l)}\mathbf{x}\!+\!\mathbf{Q}^{(l)^{\Hpow}} \mathbf{n}\!=\!\mathbf{R}^{(l)} \tilde{\mathbf{x}}^{(l)}\!+\!\tilde{\mathbf{n}}^{(l)}.
\end{equation}
The euclidean distance is thus $d(\mathbf{x})=\|\tilde{\mathbf{y}}^{(l)}-\mathbf{R}^{(l)} \tilde{\mathbf{x}}^{(l)}\|^{2}$.
For each iteration $l$, following pre-processing, the root-layer symbols, $\tilde{x}_{Q_{\mathrm{t}}}$ of $\tilde{\mathbf{x}}^{(l)}$, are exhaustively searched and successively projected to obtain a candidate vector, $\hat{\mbf{x}}$, to be accumulated in a set of searched vectors, $\mathcal{S}^{\mathrm{LORD}(l)}$, of cardinality $\abs{\mathcal{X}}$. Hence,
\begin{equation}
\label{eq:x_lord}
\hat{x}_{q_{\mathrm{t}}}^{(l)}\!\left(\tilde{x}_{Q_{\mathrm{t}}}\right)\!=\!\left\{\!\!\begin{array}{l}
\tilde{x}_{Q_{\mathrm{t}}}, \quad q_{\mathrm{t}}=Q_{\mathrm{t}} \\
\underset{x \in \mathcal{X}}{\arg \min }\left|\tilde{y}_{q_{\mathrm{t}}}^{(l)}\!-\!r_{q_{\mathrm{t}}, q_{\mathrm{t}}}^{(l)} x\!-\!\sum_{e=q_{\mathrm{t}}+1}^{Q_{\mathrm{t}}} r_{q_{\mathrm{t}}, e}^{(l)} \hat{x}_e^{(l)}\right|, q_{\mathrm{t}}\!<\!Q_{\mathrm{t}}.
\end{array}\right.
\end{equation}
The $Q_{\mathrm{t}}$ $\mathcal{S}^{\mathrm{LORD}}$s form a reduced search set for \gls{so} computations, and the \gls{llr}s are computed as
\begin{equation}
\label{eq:LLR_LORD}
    \gamma_{i,q_{\mathrm{t}}}^{\mathrm{LORD}(l)}\!\! =\!\! \frac{1}{\sigma^2}\! \! \min_{\mathbf{x} \in \mathcal{S}_{i,q_{\mathrm{t}}}^{ \mathrm{LORD}(l,1)}}\!\!\|\tilde{\mathbf{y}}^{(l)}\!-\!\mathbf{R}^{(l)} \mathbf{x}\|_{2}^{2}-\!\!\!\min _{\mathbf{x} \in \mathcal{S}_{i,q_{\mathrm{t}}}^{\mathrm{LORD}(l,0)}}\!\!\|\tilde{\mathbf{y}}^{(l)}\!-\!\mathbf{R}^{(l)} \mathbf{x}\|_{2}^{2},
\end{equation}
where $\mathcal{S}_{i,q_{\mathrm{t}}}^{\mathrm{LORD}(l,0)}\!=\!\{\Pi^{(l)} \hat{\mathbf{x}} \!\in\! \mathcal{S}^{\mathrm{LORD}(l)}\!: \hat{x}_{i,q_{\mathrm{t}}}\!=\!0\}$, $\mathcal{S}_{i,q_{\mathrm{t}}}^{\mathrm{LORD}(l,1)}\!=\!\{\Pi^{(l)} \hat{\mathbf{x}} \!\in\! \mathcal{S}^{\mathrm{LORD}(l)}\!: \hat{x}_{i,q_{\mathrm{t}}}\!=\!1\}$. To enhance performance, a \gls{v-lord} adds a post-processing step \cite{Sarieddeen8186206}, by choosing the output with the minimum distance $d(\mbf{x})$ among the best ${\mbf{x}^{\mathrm{LORD}(l)}}$ candidates per decomposition. This global distance accumulation is valid since permuted QRDs preserve the search vector space.

\subsection{Channel-punctured subspace detectors}

Channel puncturing restructures channel matrices for enhanced parallelizability, reducing detection complexity. \gls{qrd} can be replaced with a generalized \gls{wrd} \cite{Sarieddeen8186206}, $\mbf{H} \!=\! \mbf{W} \dot{\mbf{R}}$, transforming $\mbf{H}$ into a punctured upper triangular matrix $\dot{\mbf{R}} \!\in\! \mathbb{C}^{Q_{\mathrm{t}} \times Q_{\mathrm{t}}}$, where $\mbf{W} \!\in\! \mathbb{C}^{Q_{\mathrm{r}} \times Q_{\mathrm{t}}}$ is not unitary. WRD is derived from \gls{qrd} via elementary matrix operations \cite{Sarieddeen8186206}, puncturing $\mbf{R}$ entries between the diagonal and the last column, yielding the baseband input-output relation,
\begin{equation}
\Dot{\tilde{\mathbf{y}}}=\mathbf{W}^{\Hpow} \mathbf{y}=\dot{\mbf{R}} \mathbf{x}+\mathbf{W}^{\Hpow} \mathbf{n}.
\label{eq:y_puncturing}    
\end{equation}
A typical SSD routine cyclically shifts the columns of $\mbf{H}$ and generates the corresponding punctured $\dot{\mbf{R}}$ for each of the $Q_{\mathrm{t}}$ iterations, as shown in Fig. \ref{fig:aosa_tx_rx}. The \gls{llr}s of the $l^{\mathrm{th}}$ symbol bits are generated from the $l^{\mathrm{th}}$ WRD as
\begin{equation}
\label{eq:LLR_SSD}
\begin{aligned}
    \gamma_{i,q_{\mathrm{t}}}^{\mathrm{SSD}(l)}\!&=\!\frac{1}{{\sigma_{i}}^2} \times \\
    &
    \left(\min _{\mathbf{x} \in \mathcal{S}_{i,q_{\mathrm{t}}}^{ \mathrm{SSD}(l,1)}}\!\|\tilde{\mathbf{y}}^{(l)}\!-\!\dot{\mbf{R}}^{(l)} \mathbf{x}\|_{2}^{2}  \!-\!\min _{\mathbf{x} \in \mathcal{S}_{i,q_{\mathrm{t}}}^{ \mathrm{SSD}(l,0)}}\!\|\tilde{\mathbf{y}}^{(l)}\!-\!\dot{\mbf{R}}^{(l)} \mathbf{x}\|_{2}^{2}\right),
\end{aligned}
\end{equation}
where $\mathcal{S}_{i,q_{\mathrm{t}}}^{ \mathrm{SSD}(l,1)}$ and $\mathcal{S}_{i,q_{\mathrm{t}}}^{ \mathrm{SSD}(l,0)}$ are similarly defined over WRD-decomposed channels. Unlike the unitary multiplication with $\mathbf{Q}^{(l)^{\Hpow}}$ in \eqref{qld_sys}, multiplying by $\mathbf{W}^{{\Hpow}}$ in \eqref{eq:y_puncturing} results in a scaled noise variance. Since $\mathbf{W}$ is non-unitary, distances from different decompositions belong to different vector spaces, limiting the benefits of global distance post-processing (as in LORD). However, this independence across decompositions enables fully parallelized \gls{llr} computations per symbol, scaling with the \gls{mimo} dimension (see Fig. \ref{fig:aosa_tx_rx}). This parallelism can help achieve \gls{tbps} rates in \gls{thz} transceivers, as we will further illustrate in Sec. \ref{sec:complexity}. Although \gls{ssd} typically incurs a performance loss due to the non-unitary $\mathbf{W}$, puncturing in ill-conditioned \gls{thz} channels can reduce channel correlation, potentially improving performance \cite{JemaaDetection2022}, as clarified in Sec. \ref{sec:sim_res_disc}.

\subsection{Sorted-QR layered detector (SQLD)}

Among enhancements to \gls{qrd}-based detectors, \gls{sqrd} stands out by improving detection through column reordering of $\mathbf{H}$ before decomposition. This low-complexity reordering prioritizes symbols with stronger channel gains, enhancing the conditioning of $\mathbf{R}$, reducing noise amplification and interference, and lowering detection errors \cite{kobayashi2016}. The reordered received signal is then transformed using $\mathbf{Q}^{\Hpow}$, with successive interference cancellation \gls{zf} detection performed similarly to the unsorted \gls{qrd} case as in \eqref{eq:x_lord}. While previous works have combined sorting with \gls{lord} to enhance performance, they typically increase complexity. For instance, \cite{shen2007generalized} proposed \gls{g-lord} for golden code detection by expanding the lattice search to balance error performance and complexity, but its focus on golden code limits generalizability and increases complexity in high-dimensional systems. Similarly, \gls{plr} \cite{izadinasab2019partial} embeds sorting to improve efficiency by reducing selected subchannels in large-scale \gls{mimo}, but its reliance on quasi-static channels limits applicability to dynamic environments like \gls{thz} systems.

To address the challenges of ill-conditioned THz channels and the need for high parallelizability, we explore a combination of \gls{lord}, sorting, and subspace puncturing in Algorithm \ref{alg:SQLD}. The algorithm begins by sorting the columns of the channel matrix $\mathbf{H}$ using a chosen sorting criterion, resulting in a permutation matrix $\mathbf{\Pi}_{S}$ and a corresponding permutation vector $\mathbf{p}_{S}$ that indicates the new column order. The sorting criterion can be selected based on different objectives, such as maximizing the channel column magnitudes or optimizing the signal-to-noise ratio (SNR), as in \gls{v-blast} \cite{wolniansky1998v}. This ordering enhances the \gls{sic} performance, which is repeated in every \gls{cd} across different permutations. After sorting, each permutation step is performed independently for each layer \( l = 1, 2, \dots, Q_{\mathrm{t}} \). In each step, a permutation matrix \( \mathbf{P}_{l} \) is constructed to swap the column initially in the \( i \)-th position of \( \mathbf{H} \) (now located at \( \mbf{p}_{S}(i) \) after sorting) with the last column. This produces the permuted channel matrix \( \mathbf{H}^{(l)} \), derived by applying the cumulative permutation \( \mathbf{\Pi}^{(l)} = \mathbf{\Pi}_{S} \mathbf{P}_{l} \) to \( \mathbf{H} \). 
\begin{algorithm}[t!]
\caption{Sorted-QR layered detector (SQLD)}
\label{alg:SQLD}
\begin{algorithmic}[1]
\Require $\mathbf{y}$, $\mathbf{H}$, $\mathcal{X}$, $\sigma$, $\eta$
\Ensure $\hat{\mathbf{x}}^{\mathrm{SQLD}}$,  $\mathbf{\gamma}^{\mathrm{SQLD}}$

\State Sort $\mathbf{H}$ according to \cite{wolniansky1998v}; obtain $\mathbf{\Pi}_{S}$ and $\mathbf{p}_{S}$

\For{$l = 1$ to $Q_{\mathrm{t}}$} \Comment{Can be parallelized}
    \State Construct permutation matrix $\mathbf{P}_{l}$  
    \State $\mathbf{H}^{(l)} \gets \mathbf{H} \mathbf{\Pi}^{(l)}$, where $\mathbf{\Pi}^{(l)} = \mathbf{\Pi}_{S} \times \mathbf{P}_{l}$
    \If{$\eta \leq l$}
   
        \State $\mathbf{H}^{(l)} = \mathbf{W}^{(l)} \dot{\mathbf{R}}^{(l)}$\Comment{WRD}
        \State $\tilde{\mathbf{y}}^{(l)} \gets {\mathbf{W}^{(l)}}^{\mathrm{H}} \mathbf{y}$ 
    \Else
        \State $\mathbf{H}^{(l)} = \mathbf{Q}^{(l)} \mathbf{R}^{(l)}$\Comment{QRD}
        \State $\tilde{\mathbf{y}}^{(l)} \gets {\mathbf{Q}^{(l)}}^{\mathrm{H}} \mathbf{y}$ 
    \EndIf

    \ForAll{$\tilde{x}_{Q_{\mathrm{t}}} \in \mathcal{X}$}  \Comment{Over root-layer symbols}
        \State $\hat{x}_{q_{\mathrm{t}}}^{(l)} \gets$ ~\eqref{eq:x_lord} for $q_{\mathrm{t}} \in [1 \dots Q_{\mathrm{t}}]$
        \If{$\eta \leq l$}
            \State $\gamma_{i,l} \gets$ ~\eqref{eq:LLR_SSD} for $i \in [1 \dots \log_2{|\mathcal{X}|}]$
        \Else
            \State $\gamma_{i,l} \gets$ ~\eqref{eq:LLR_LORD} for $i \in [1 \dots \log_2{|\mathcal{X}|}]$
        \EndIf
    \EndFor
\EndFor

\end{algorithmic}
\end{algorithm}
Following sorting/permutation, the subspace detection procedure is performed for each layer \( l \). We propose an enhanced subspace procedure inspired by \cite{JemaaDetection2022}, where puncturing mitigates performance degradation in \gls{thz} channels. A key parameter, \( \eta \in \{1, 2, \dots, Q_{\mathrm{t}}\} \), specifies the number of layers subjected to puncturing and, consequently, the number of non-punctured layers that undergo exhaustive search. Therefore, Algorithm \ref{alg:SQLD} integrates an initial sorting step and embeds puncturing to improve parallelizability while fine-tuning \( \eta \) to balance performance and complexity. By exhaustively searching the \( Q_{\mathrm{t}} - \eta \) most reliable (after sorting) lower layers, we prioritize accuracy in the initial stages, which is crucial for mitigating error propagation. This is complemented by WRD-based projections at higher layers, leveraging the complexity and performance advantages of \gls{ssd} over \gls{lord}.

The \gls{sqld} approach reduces complexity, enhances parallelizability through WRD, and improves performance without increasing the search list size, which would be impractical under \gls{thz} \gls{um-mimo} constraints. The puncturing techniques can be customized based on design needs; alternative puncturing patterns beyond the conventional V-shape \cite{mansour2020low} provide greater flexibility for performance and complexity tuning.

\section{Performance Analysis}
\label{sec:MIMO_perf_analysis}
Analyzing the performance of \gls{thz} \gls{mimo} systems is particularly challenging because of the intricate and often correlated channel structures combined with \gls{thz}-specific impairments such as severe path loss, molecular absorption, and hardware constraints. These factors not only increase the complexity of accurately modeling the distribution of \gls{mimo} channels but also significantly limit analytical tractability. As a result, obtaining closed-form expressions or rigorous performance bounds typically requires introducing substantial assumptions and mathematical simplifications. In addition, the \gls{thz} channel models and the correlation structure itself are not universal but rather scenario-specific. They depend strongly on the channel conditions, the type of communication scenario, and the system dimensionality. Certain correlation models fail to capture the behavior of high-dimensional systems, while others are purely statistical in nature and therefore not analytically tractable, which further complicates performance analysis in the \gls{thz} context.

\subsection{Pairwise error probability for ML and PML detectors}
Our analysis here focuses on the \gls{pep} and diversity under conventional and parallelizable (channel-punctured) schemes, assuming optimal detection in both cases. We compare the optimal ML detector to the \gls{pml} detector, which performs the exhaustive search in \eqref{ML_equation} over a punctured channel, where
\begin{equation}
\mathbf{\hat{x}}^{\mathrm{PML}}\!=\!\underset{\mathbf{x} \in {\mathcal{\bar{X}}}}{\arg \min }\| \mathbf{W^{\Hpow}}(\mathbf{y}\!-\!\mathbf{H x})\|_{2}^{2}\!=\!\underset{\mathbf{x} \in {\mathcal{\bar{X}}}}{\arg \min }\|\tilde{\mathbf{y}}\!-\!\dot{\mbf{R}} \mathbf{x}\|_{2}^{2}.
\end{equation}
The \gls{pep} for \gls{pml}, $\mathrm{Pr}_{\mathrm{P}}^{\mathrm{PML}}$, which represents the probability of transmitting $\mathbf{x}^{(1)}$ and detecting $\mathbf{x}^{(2)} \neq \mathbf{x}^{(1)}$, is expressed as
\begin{equation}
\begin{aligned}
    \mathrm{Pr}_{\mathrm{P}}^{\mathrm{PML}}
    & =\mathrm{Pr}\left(\left\|\mathbf{W}^{\Hpow}\left(\mathbf{y}-\mathbf{H} \mathbf{x}^{(2)}\right)\right\|_{2}^2 \leq\left\|\mathbf{W}^{\Hpow}\left(\mathbf{y}-\mathbf{H} \mathbf{x}^{(1)}\right)\right\|_{2}^2\right) . 
\end{aligned}
\end{equation}
When $\mathbf{x}^{(1)}$ is transmitted, $\mathbf{y}\!=\!\mathbf{H} \mathbf{x}^{(1)}\!+\!\mathbf{n}$, and
\begin{equation}
\label{eq:PML_1}
\begin{aligned}
    \mathrm{Pr}_{\mathrm{P}}^{\mathrm{PML}}
    & =\mathrm{Pr}\left(\left\|\mathbf{W}^{\Hpow}\left(\mathbf{n}-\mathbf{H} \mathbf{d}\right)\right\|_{2}^2 \leq\left\|\mathbf{W}^{\Hpow}\mathbf{n}\right\|_{2}^2\right) \\
    & =\mathrm{Pr}\left(\Re\left(\mathbf{n}^{\Hpow} \mathbf{W} \dot{\mbf{R}} \mathbf{d}\right) \geq \frac{1}{2}\|\dot{\mbf{R}} \mathbf{d}\|_{2}^2\right), 
\end{aligned}
\end{equation}
where $ \mathbf{d}=\mathbf{x}^{(2)} -\mathbf{x}^{(1)}$. Conditioning on $\mathbf{W}$ and $\dot{\mbf{R}}$, $\Re{(\mathbf{n}^{\Hpow} \mathbf{W}\dot{\mbf{R}}\mathbf{d})}$ is a Gaussian random variable of mean and variance,
\begin{align}
\mathbb{E}\!\left[\mathbf{n}^{\Hpow} \mathbf{W} \dot{\mathbf{R}} \mathbf{d}\right] 
&= \mathbb{E}\!\left[\mathbf{n}^{\Hpow} \right] \mathbf{W} \dot{\mathbf{R}} \mathbf{d} \stackrel{(\text{a})}{=} 0, \notag \\
\mathbb{E}\!\left[\!\left(\mathbf{n}^{\Hpow} \mathbf{W} \dot{\mathbf{R}} \mathbf{d}\right)\!\left(\mathbf{n}^{\Hpow} \mathbf{W} \dot{\mathbf{R}} \mathbf{d}\right)^{\Hpow}\!\right] 
&\stackrel{(\text{b})}{=} \mathbb{E}\!\left[\mathrm{Tr}\!\left(\!\left(\mathbf{n}^{\Hpow} \mathbf{W} \dot{\mathbf{R}} \mathbf{d}\right)\!\left(\mathbf{n}^{\Hpow} \mathbf{W} \dot{\mathbf{R}} \mathbf{d}\right)^{\Hpow}\!\right)\!\right] \notag \\
&\stackrel{(\text{c})}{=} \mathbb{E}\!\left[\mathrm{Tr}\!\left(\!\left( \mathbf{W} \dot{\mathbf{R}} \mathbf{d} \right)^{\Hpow} \mathbf{n} \mathbf{n}^{\Hpow} \left( \mathbf{W} \dot{\mathbf{R}} \mathbf{d} \right)\!\right)\!\right] \notag \\
&\stackrel{(\text{d})}{=} \sigma^2 \left\|\mathbf{W} \dot{\mathbf{R}} \mathbf{d}\right\|_2^2.
\label{eq:PML_2}
\end{align}
where $\stackrel{(\text{a})}{=}$ holds because $\mathbb{E}\left[\mathbf{n}\right] \!=\! 0$, $\stackrel{(\text{b})}{=}$ because $\mathbf{n}^{\Hpow} \mathbf{W} \dot{\mbf{R}} \mathbf{d}$ is a scalar and thus the trace operator can be exchanged with the expectation, $\stackrel{(\text{c})}{=}$ since $\mathrm{Tr}(\mathbf{AB})\!=\!\mathrm{Tr}(\mathbf{BA})$, and $\stackrel{(\text{d})}{=}$ because $\mathbb{E}\left[\mathbf{n}\mathbf{n}^{\Hpow}\right] \!=\! \sigma^2 \mathbf{I}_{Q_{\mathrm{r}}}$.
To simplify the expression and highlight the characteristics of puncturing, we derive the upper bound using the properties of a Gaussian-distributed random variable, where the probability of exceeding a given distance is expressed through the $Q$ function as 
\begin{equation}
    \begin{aligned}
    \mathrm{Pr}_{\mathrm{P}}^{\mathrm{PML}} 
    & =\mathbb{E}\left[Q\left(\frac{\norm{\dot{\mathbf{R}}\mathbf{d}}_{2}^2}{2\sigma \norm{ \mathbf{W} \dot{\mathbf{R}} \mathbf{d}}_{2}}\right)\right] \stackrel{(\text{a})}{\le} \mathbb{E}\left[Q\left(\frac{\norm{\dot{\mathbf{R}}\mathbf{d}}_{2}}{2\sigma \norm{ \mathbf{W}}_{2}}\right)\right] \\
    & \stackrel{(\text{b})}{\le} \mathbb{E}\left[Q\left(\frac{\norm{\dot{\mathbf{R}}\mathbf{d}}_{2}}{2\sigma \norm{ \mathbf{W}}_{\mathrm{F}}}\right)\right]  \stackrel{(\text{c})}{=}\mathbb{E}\left[Q\left(\frac{\norm{\dot{\mathbf{R}}\mathbf{d}}_{2}}{2\sigma \sqrt{Q_{\mathrm{r}}}}\right)\right] \\
    & 
    \stackrel{(\text{d})}{\approx} \mathbb{E}\left[
    \frac{e^{-\frac{\norm{\dot{\mathbf{R}}\mathbf{d}}_{2}^{2}}{8\sigma^2 Q_{\mathrm{r}}}}}{12}\right]+ \mathbb{E}\left[\frac{e^{-\frac{\norm{\dot{\mathbf{R}}\mathbf{d}}_{2}^{2}}{12\sigma^2 Q_{\mathrm{r}}}}}{4}\right],
    \label{eq:Prp_PML}
    \end{aligned}
\end{equation}
where $\stackrel{(\text{a})}{\le}$ holds because $\norm{\mathbf{AB}}_{2} \!\le\! \norm{\mathbf{A}}_{2} \norm{\mathbf{B}}_{2}$ and $\stackrel{(\text{b})}{\le}$ holds because $\norm{\mathbf{A}}_{2} \!\le\! \norm{\mathbf{A}}_{\mathrm{F}}$ and the $Q$ function is monotonically decreasing in $\mathbb{R}^{+}$. By construction, the columns of $\mathbf{W}$ are normalized, hence, $\norm{\mathbf{W}}_{\mathrm{F}}\!=\!\sqrt{Q_{\mathrm{r}}}$ is leveraged in $\stackrel{(\text{c})}{=}$. Because obtaining the distributions of $\mathbf{\dot{R}}$ and $\mathbf{W}$ is challenging, and introducing the $Q\mhyphen$function further complicates analytical tractability, we can upper bound the $Q\mhyphen$function using $Q(x)\!\le\! e^{-\frac{x^2}{2}}$ or one of its approximations. The approximation in $\stackrel{(\text{d})}{\approx}$ follows $Q(x) \!\approx\! \frac{e^{-\frac{x^2}{2}}}{12}\!+\!\frac{e^{-\frac{2 x^2}{3}}}{4}$ \cite{1210748}. 

For regular ML detection, the \gls{pep}, $\mathrm{Pr}_{\mathrm{P}}^{\mathrm{ML}}$, can be derived using similar steps. However, the unitary behavior of $\mathbf{Q}$ results in $\norm{\mathbf{H}\mathbf{d}}_{2}= \norm{\mathbf{Q}\mathbf{R}\mathbf{d}}_{2} =\norm{\mathbf{R}\mathbf{d}}_{2}$. Therefore,
\begin{equation}
    \begin{aligned}
    \mathrm{Pr}_{\mathrm{P}}^{\mathrm{ML}}
    & =\mathbb{E}\left[Q\left(\frac{\norm{{\mathbf{H}}\mathbf{d}}_{2}}{2\sigma}\right)\right]=\mathbb{E}\left[Q\left(\frac{\norm{{\mathbf{R}}\mathbf{d}}_{2}}{2\sigma}\right)\right] \\
    & \stackrel{(\text{h})}{\approx} \mathbb{E}\left[
    \frac{e^{-\frac{\norm{{\mathbf{R}}\mathbf{d}}_{2}^{2}}{8\sigma^2 }}}{12}\right]+ \mathbb{E}\left[\frac{e^{-\frac{\norm{{\mathbf{R}}\mathbf{d}}_{2}^{2}}{12\sigma^2 }}}{4}\right].
    \end{aligned} 
    \label{eq:Prp_ML}
\end{equation}

\subsection{Closed-form \texorpdfstring{$\mathrm{Pr}_{\mathrm{p}}$}\ \ lower bound for \texorpdfstring{$\alpha\mhyphen\mu$}\ -based indoor THz channels}

We start from \eqref{eq:Prp_ML}, and note that $\norm{\mathbf{Hd}}_{2} = \norm{\mathbf{Hd}}_{\mathrm{F}}\!\le\! \norm{\mathbf{H}}_{\mathrm{F}} \norm{\mathbf{d}}_{\mathrm{F}} = \norm{\mathbf{H}}_{\mathrm{F}} \norm{\mathbf{d}}_{2}$. Given that $Q(\cdot)$ is a monotonically decreasing function, this inequality allows us to establish a lower bound on $\mathrm{Pr}_{\mathrm{P}}^{\mathrm{ML}}$ as
\begin{equation}
    \begin{aligned}
        \mathrm{Pr}_{\mathrm{P}}^{\mathrm{ML}} & \geq \mathbb{E}\left[Q\left(\frac{\norm{\mathbf{H}}_{\mathrm{F}} \norm{\mathbf{d}}_{2}}{2\sigma}\right)\right] = \mathbb{E}\left[Q\left(\sqrt{\frac{\norm{\mathbf{H}}_{\mathrm{F}}^{2} \norm{\mathbf{d}}_{2}^{2}}{4\sigma^{2}}}\right)\right] \\
        & = \int_0^{\infty} Q\left(\frac{\norm{\mathbf{d}}_{2}}{2\sigma} \sqrt{z}\right) f_{\norm{\mathbf{H}}_{\mathrm{F}}^{2}}(z) dz.
    \end{aligned}
    \label{eq:lower_bound_ML_1}
\end{equation}

To obtain a closed-form lower bound, we need to evaluate \eqref{eq:lower_bound_ML_1} after deriving the \gls{pdf} of $\norm{\mathbf{H}}_{\mathrm{F}}^{2} = \sum_{q_{\mathrm{r}}=1}^{Q_{\mathrm{r}}} \sum_{q_{\mathrm{t}}=1}^{Q_{\mathrm{t}}}|h_{q_{\mathrm{r}},q_{\mathrm{t}}}|^2$. To derive a THz-specific bound, we employ the $\alpha\mhyphen\mu$ \gls{thz} channel model, which accurately characterizes \gls{thz} indoor propagation conditions \cite{Jemaa2024Performance,payami2020accurate}. The channel entries $h_{q_{\mathrm{r}},q_{\mathrm{t}}}$ values are assumed to be either independent or follow a structured correlation model, such as the exponential correlation. Given the lack of measurement-verified indoor \gls{thz} channel correlation models, we consider both cases in order to assess the impact of correlation together with \gls{thz}-specific path loss and small-scale fading characteristics. Under the exponential correlation model, the channel matrix can be expressed as  
\begin{align}
    \mathbf{H}_{\mathrm{corr}} = \mathbf{R}_{\mathrm{r}}^{1/2} \, \mathbf{H} \, \mathbf{R}_{\mathrm{t}}^{1/2},
\end{align}
where the correlation matrices are defined by ${\mathbf{R}_{\mathrm{r}}}_{i,j} = \rho_{\mathrm{r}}^{|i - j|}$ and ${\mathbf{R}_{\mathrm{t}}}_{i,j} = \rho_{\mathrm{t}}^{|i - j|}$, with correlation coefficients $0 \leq \rho_{\mathrm{r}}, \rho_{\mathrm{t}} \leq 1$. This formulation reflects the widely used Kronecker-type exponential correlation structure. Because $\mathbf{R}_r$ and $\mathbf{R}_t$ are deterministic, and  
\begin{align}
    \norm{\mathbf{H}_{\mathrm{corr}}}_{\mathrm{F}} \le \norm{\mathbf{R}_r^{1/2}}_{\mathrm{F}} \, \norm{\mathbf{H}}_{\mathrm{F}} \, \norm{\mathbf{R}_t^{1/2}}_{\mathrm{F}},
\end{align}
applying the Frobenius norm inequality for matrix products reduces the problem to deriving the \gls{pdf} of $\norm{\mathbf{H}}_{\mathrm{F}}^2$ under independent channel entries. In this way, the correlation influence is absorbed into fixed scaling factors from $\mathbf{R}_{\mathrm{r}}$ and $\mathbf{R}_{\mathrm{t}}$, while the randomness is fully determined by the distribution of the independent entries of $\mathbf{H}$. Each entry of the channel $\mathbf{H}$ is modeled as $|h_{q_{\mathrm{r}},q_{\mathrm{t}}}| \!=\! |h_{\mathrm{p}_{q_{\mathrm{r}},q_{\mathrm{t}}}}| |h_{\mathrm{f}_{q_{\mathrm{r}},q_{\mathrm{t}}}}|$, where $h_{\mathrm{p}_{q_{\mathrm{r}},q_{\mathrm{t}}}}$ represents large-scale path loss and $h_{\mathrm{f}_{q_{\mathrm{r}},q_{\mathrm{t}}}}$ models $\alpha\mhyphen\mu$ small-scale fading caused by multipath components.

\begin{theorem}
\label{thm:pdf_alpha}
Let $X$ be an $\alpha\mhyphen\mu$-distributed random variable and $c\!>\!0$ be a constant. Then, $(cX)^2$ is also an $\alpha\mhyphen\mu$-distributed random variable.
\end{theorem}

\begin{proof}
Given an $\alpha\mhyphen\mu$-distributed random variable $X$, where $\mathbb{E}[X] \!=\! \bar{x}$ and $\beta \!=\! \frac{\Gamma\left(\mu+\frac{1}{\alpha}\right)}{\Gamma(\mu)}$, and a constant $c\!>\!0$, we have $f_{cX}(x) = \frac{1}{c} f_X\left(\frac{x}{c}\right)=\frac{\alpha \beta^{\alpha \mu} x^{\alpha \mu-1}}{\bar{x}^{\alpha \mu} c^{\alpha \mu} \Gamma(\mu)}e^{-\left(\frac{\beta x}{c\bar{x}}\right)^\alpha}$, and $cX$ is also $\alpha\mhyphen\mu$-distributed with the same parameters, but with a modified mean $\mathbb{E}[cX] \!=\! c \bar{x}$. Furthermore, the \gls{pdf} of $Y = X^2$ can be expressed as 
\begin{equation}
\begin{aligned}
    f_{Y}(x) &= \frac{1}{2\sqrt{x}} f_X\left(\sqrt{x}\right) = \frac{\alpha \beta^{\alpha \mu} x^{\frac{\alpha}{2} \mu-1}}{2\bar{x}^{\alpha \mu} \Gamma(\mu)}
    e^{\left(-\left(\frac{\beta^2 x}{ \bar{x}^2}\right)^{\frac{\alpha}{2}} \right)}. 
\end{aligned}
\end{equation}
The $k^{\mathrm{th}}$ moment for an $\alpha\mhyphen\mu$ random variable follows directly from its \gls{pdf} and is readily obtained as $\mathbb{E}\left[X^k\right] \!=\! \left(\frac{\bar{x}}{\beta}\right)^k \frac{\Gamma\left(\mu+\frac{k}{\alpha}\right)}{\Gamma(\mu)}$. Thus, $\mathbb{E}[Y] \!=\! \mathbb{E}\left[X^2\right] \!=\! \left(\frac{\bar{x}}{\beta}\right)^2 \frac{\Gamma\left(\mu+\frac{2}{\alpha}\right)}{\Gamma(\mu)} = \bar{y}$. This implies $\left(\frac{\beta}{\bar{x}}\right)^2 \!=\! \frac{\Gamma\left(\mu+\frac{2}{\alpha}\right)}{\Gamma(\mu) \bar{y}}$. It follows that
\begin{equation}
\begin{aligned}
    f_Y(y) &= \frac{\frac{\alpha}{2} \left(\frac{\beta^2}{x^2}\right)^{\frac{\alpha}{2}\mu}}{\Gamma(\mu)}
    y^{\frac{\alpha}{2} \mu - 1} 
    e^{- \left(\frac{\beta_y}{\bar{y}} y\right)^{\alpha / 2}} \\
    & = \frac{\frac{\alpha}{2} \beta_y^{\frac{\alpha}{2}\mu}}{\bar{y}^{\frac{\alpha}{2}\mu}\Gamma(\mu)}
    y^{\frac{\alpha}{2} \mu - 1} 
    e^{- \left(\frac{\beta_y}{\bar{y}} y\right)^{\alpha / 2}},
\end{aligned}
\end{equation}
meaning $Y$ is also $\alpha\mhyphen\mu$-distributed, with $\alpha_Y \!=\! \frac{\alpha}{2}$, $\mu_Y \!=\! \mu$, and $\beta_Y \!=\! \frac{\Gamma\left(\mu+\frac{2}{\alpha}\right)}{\Gamma(\mu)}$. Therefore, if $cX\!\sim\! \alpha\mhyphen\mu$, $(cX)^2\!\sim\! \alpha\mhyphen\mu$, which concludes the proof.
\end{proof}

Using theorem \ref{thm:pdf_alpha}, for $\left|h_{\mathrm{f}_{q_{\mathrm{r}},q_{\mathrm{t}}}}\right| \!\sim\! \alpha\mhyphen\mu$ and a constant path loss $ h_{\mathrm{p}_{q_{\mathrm{r}},q_{\mathrm{t}}}}$, $|h|_{q_{\mathrm{r}},q_{\mathrm{t}}}^2 \!=\! \left| h_{\mathrm{p}_{q_{\mathrm{r}},q_{\mathrm{t}}}} h_{\mathrm{f}_{q_{\mathrm{r}},q_{\mathrm{t}}}} \right|^2\!\sim\! \alpha\mhyphen\mu$ with $\alpha \!=\! \frac{\alpha}{2}$, $\mu$, and $\beta \!=\! \frac{\Gamma\left(\mu+\frac{2}{\alpha}\right)}{\Gamma(\mu)}$.

As stated, our goal is to derive the distribution of $\norm{\mathbf{H}}_{\mathrm{F}}^2$. We extend the approach in \cite{payami2020accurate}, which approximates the sum of $\alpha\mhyphen\mu$ random variables using Puiseux series expansion and moment matching. More specifically, let 
\begin{equation}
    Z = \norm{\mathbf{H}}_{\mathrm{F}}^2 = \sum_{q_{\mathrm{r}}=1}^{Q_{\mathrm{r}}} \sum_{q_{\mathrm{t}}=1}^{Q_{\mathrm{t}}}|h_{q_{\mathrm{r}},q_{\mathrm{t}}}|^2 = \sum_{q_{\mathrm{r}}=1}^{Q_{\mathrm{r}}} \sum_{q_{\mathrm{t}}=1}^{Q_{\mathrm{t}}} Y_{q_{\mathrm{r}},q_{\mathrm{t}}}.
\end{equation}
By applying \cite[Theorem 1]{payami2020accurate} for $L \!=\!{Q_{\mathrm{r}}Q_{\mathrm{t}}}$, $\alpha_{\mathrm{z}} \!=\! \frac{\alpha}{2}$, $\mu_{\mathrm{z}} \!=\! \sum_{q_{\mathrm{r}}=1}^{Q_{\mathrm{r}}} \sum_{q_{\mathrm{t}}=1}^{Q_{\mathrm{t}}} \mu_{q_{\mathrm{r}},q_{\mathrm{t}}}$, $\bar{Z} \!=\! \sum_{q_{\mathrm{r}}=1}^{Q_{\mathrm{r}}} \sum_{q_{\mathrm{t}}=1}^{Q_{\mathrm{t}}} |\bar{h}_{q_{\mathrm{r}},q_{\mathrm{t}}}|^2$, and $\beta_z \!=\! \frac{\Gamma\left(\mu_z + \frac{1}{\alpha_z}\right)}{\Gamma(\mu_z)}$, the \gls{pdf} of $Z = \norm{\mathbf{H}}_{\mathrm{F}}^2$ can be approximated as
\begin{equation}
\begin{aligned}
    f_Z(z) &\approx \sum_{m=1}^{\Psi} 
    \frac{c_m \alpha_z \beta_z^{\alpha_z \mu_z} z^{\alpha_z \mu_z - 1}}{(\omega_m \bar{z})^{\alpha_z \mu_z} \Gamma(\mu_z)}
    e^{-\left(\beta_z \frac{z}{\omega_m \bar{z}}\right)^{\alpha_z}},
    \label{eq:alpha_mu_frob}
\end{aligned}
\end{equation}
for any arbitrary integer $\Psi$. The parameters $c_m$ and $\omega_m$ are determined by solving the system of linear equations \cite{payami2020accurate},
\begin{equation}
\begin{aligned}
    \sum_{m=1}^{\Psi} c_m \omega_m^n &= \frac{\mathbb{E}\left[Z^n\right]}{\mathbb{E}^n\left[Z\right]} \xi^{(n)}, \quad n = 0, 1, 2, \ldots, 2M-2, \\
    \sum_{m=1}^{\Psi} \frac{c_m}{\omega_m^{\alpha_{\mathrm{z}} \mu_z}} &= \alpha_{\mathrm{z}}^{L-1} 
    \frac{\bar{z}^{\alpha_{\mathrm{z}} \mu_z} \Gamma(\mu_z)}{\beta_z^{\alpha_{\mathrm{z}} \mu_z} \Gamma(\alpha_{\mathrm{z}} \mu_z)} \quad \times \prod_{i=1}^{L} \frac{\beta_i^{\alpha_{\mathrm{z}} \mu_i} \Gamma(\alpha_{\mathrm{z}} \mu_i)}{\bar{y}_i^{\alpha_{\mathrm{z}} \mu_i} \Gamma(\mu_i)}.
\end{aligned}
\label{eq:moment_matching_sys}
\end{equation}
where $\xi^{(n)} = \frac{\Gamma\left(\mu_z + \frac{n}{\alpha_{\mathrm{z}}}\right) \Gamma^{n-1}(\mu_z)}{\Gamma^n\left(\mu_z + \frac{1}{\alpha_{\mathrm{z}}}\right)}$. To derive $\mathbb{E}\left[Z^n\right]$, we use the following formula that relies on the individual terms of the summation in the random variable $Z$ \cite{payami2020accurate}
\begin{equation} 
    \begin{aligned} 
    \mathbb{E}\left[Z^n\right] &= \sum_{n_1=0}^{n} \sum_{n_2=0}^{n_1} \cdots \sum_{n_{L-1}=0}^{n_{L-2}} \prod_{i=1}^{L-1} \binom{n_{i-1}}{n_i} \times \prod_{i=1}^{L} \mathbb{E}\left[Y_i^{n_{i-1} - n_i}\right].
    \end{aligned} 
\end{equation}

\begin{theorem}
For the $\alpha$–$\mu$ based indoor THz channel, the $\mathrm{Pr}_{\mathrm{P}}^{\mathrm{ML}}$ can be lower bounded using \eqref{eq:lower_bound_ML_1} together with \eqref{eq:alpha_mu_frob} as follows:
\label{thm:prp_alpha}
    \begin{equation}
        \begin{aligned}
            \mathrm{Pr}_{\mathrm{P}}^{\mathrm{ML}} & \geq \sum_{m=1}^{\Psi} \frac{c_m \alpha_z \beta_z^{\alpha_z \mu_z}}{2 \sqrt{\pi}(\omega_m \bar{z})^{\alpha_z \mu_z} \Gamma(\mu_z)} \left(\frac{8\sigma^2}{\norm{\mathbf{d}}_{2}^{2}}\right)^{\alpha_z \mu_z} {H}^{1,2}_{2,2} \\
        & \left(\frac{\left(8\beta_z\sigma^2\right)^{\alpha_z}}{\left(\omega_m\bar{z}\norm{\mathbf{d}}_{2}^{2}\right)^{\alpha_z}}\middle|
        \begin{matrix}
        (1-\alpha_z\mu_z,\alpha_z),(\frac{1}{2}-\alpha_z\mu_z,\alpha_z)\\
        (0,1),(-\alpha_z\mu_z, \alpha_z)
        \end{matrix}
        \right),
        \end{aligned}
    \end{equation}
\end{theorem}
where we involve the Meijer-G and Fox-H functions \cite{kilbas2004h} adequately. 
\begin{proof}
We start the proof by using the \gls{pdf} of $\norm{\mathbf{H}}_{\mathrm{F}}^{2}$ and \({Q}(a\sqrt{x}) \!=\! \frac{1}{2} \text{erfc} \left( \sqrt{\frac{a^2 x}{2}} \right)\)in $\stackrel{(\mathrm{a})}{=}$ and derive as follows:
\begin{align}
        \mathrm{Pr}_{\mathrm{P}}^{\mathrm{ML}} & \geq  \int_0^{\infty} Q\left(\frac{\norm{\mathbf{d}}_{2}}{2\sigma} \sqrt{z}\right) f_{\norm{\mathbf{H}}_{\mathrm{F}}^{2}}(z) d z \nonumber \\
        & \stackrel{(\mathrm{a})}{=} \int_0^{\infty} \frac{1}{2} \text{erfc} \left( \sqrt{\frac{\norm{\mathbf{d}}_{2}^{2} z}{8\sigma^2}} \right) \sum_{m=1}^{\Psi} 
        \frac{c_m \alpha_z \beta_z^{\alpha_z \mu_z}}{(\omega_m \bar{z})^{\alpha_z \mu_z} \Gamma(\mu_z)} \nonumber \\
        & \quad \times z^{\alpha_z \mu_z - 1} e^{-\left(\beta_z \frac{z}{\omega_m \bar{z}}\right)^{\alpha_z}} dz \nonumber \\
        & = \sum_{m=1}^{\Psi} \int_0^{\infty} \frac{c_m \alpha_z \beta_z^{\alpha_z \mu_z}}{2(\omega_m \bar{z})^{\alpha_z \mu_z} \Gamma(\mu_z)} z^{\alpha_z \mu_z - 1} \nonumber \\
        & \quad \times \text{erfc} \left( \sqrt{\frac{\norm{\mathbf{d}}_{2}^{2} z}{8\sigma^2}} \right) e^{-\left(\beta_z \frac{z}{\omega_m \bar{z}}\right)^{\alpha_z}} dz \nonumber \\
        & \stackrel{(\mathrm{b})}{=} \sum_{m=1}^{\Psi} \int_0^{\infty} \frac{c_m \alpha_z \beta_z^{\alpha_z \mu_z}}{2 \sqrt{\pi}(\omega_m \bar{z})^{\alpha_z \mu_z} \Gamma(\mu_z)} z^{\alpha_z \mu_z - 1} {G}^{2,0}_{1,2} \nonumber \\ 
        & \left( \frac{\norm{\mathbf{d}}_{2}^{2} z}{8\sigma^2} \middle|
        \begin{matrix} 1 \\ 0, \frac{1}{2} \end{matrix}\right) G_{0,1}^{1,0} \left[ \left(\left(\frac{\beta_z}{\omega_m \bar{z}}\right)^{\alpha_z} z^{\alpha_z}\right) \Bigg| \begin{array}{c} \sim \\ 0 \end{array} \right] dz \nonumber \\ 
        & \stackrel{(\mathrm{c})}{=} \sum_{m=1}^{\Psi} \frac{c_m \alpha_z \beta_z^{\alpha_z \mu_z}}{2 \sqrt{\pi}(\omega_m \bar{z})^{\alpha_z \mu_z} \Gamma(\mu_z)} \left(\frac{8\sigma^2}{\norm{\mathbf{d}}_{2}^{2}}\right)^{\alpha_z \mu_z} {H}^{1,2}_{2,2} \nonumber \\
        & \left(\frac{\left(8\beta_z\sigma^2\right)^{\alpha_z}}{\left(\omega_m\bar{z}\norm{\mathbf{d}}_{2}^{2}\right)^{\alpha_z}}\middle|
        \begin{matrix}
        (1-\alpha_z\mu_z,\alpha_z),(\frac{1}{2}-\alpha_z\mu_z,\alpha_z)\\
        (0,1),(-\alpha_z\mu_z, \alpha_z)
        \end{matrix}
        \right),
    \label{eq:lower_bound_ML_2}
\end{align}
where in $\stackrel{(\mathrm{b})}{=}$, we express the erfc function in its Meijer-G function form \cite{prudnikov1986integrals} as \(\text{erfc}(\sqrt{x}) \!=\! \frac{1}{\sqrt{\pi}} {G}^{2,0}_{1,2} \left( x \middle|\begin{matrix}1 \\0, \frac{1}{2}\end{matrix}\right)\), and also express the exponential function using it Meijer-G function form as  $e^{-\left(\beta_z \frac{z}{\omega_m \bar{z}}\right)^{\alpha_z}} \!=\! G_{0,1}^{1,0} \left[ \left(\left(\frac{\beta_z}{\omega_m \bar{z}}\right)^{\alpha_z} z^{\alpha_z}\right) \Bigg| \begin{array}{c} \sim \\ 0 \end{array} \right]$. Finally, we use the Meijer-G function integration property \cite{website_wolfram} in $\stackrel{(\mathrm{c})}{=}$. The above Fox $H$-function representation is valid since $\alpha_z,\mu_z,\beta_z,\sigma^2,\omega_m,\|\mathbf{d}\|^2>0$, ensuring $\Re(\alpha_z\mu_z)>0$, $r=\alpha_z>0$, 
and disjoint poles in the parameter sets, which satisfy the standard existence 
conditions of the Fox $H$-function \cite{website_wolfram}.
\end{proof}

\subsection{Empirical \texorpdfstring{$\mathrm{Pr}_{\mathrm{e}}$} \ \ bounds: ML versus PML in the THz band}
The union bound links the error probability $\mathrm{Pr}_{\mathrm{e}}$ to $\mathrm{Pr}_{\mathrm{p}}$ via
\begin{equation}
\begin{aligned}
    \mathrm{Pr}_{\mathrm{e}} & \leq \sum_{\mathbf{x}^{(i)} , \mathbf{x}^{(j)}  \in \bar{\mathcal{X}}} a(\mathbf{x}^{(i)} , \mathbf{x}^{(j)}) \mathrm{Pr}(\mathbf{x}^{(i)} ) \mathrm{Pr}_{\mathrm{P}}(\mathbf{x}^{(j)} / \mathbf{x}^{(i)}) \\
    & = \frac{1}{|\mathcal{X}|}\sum_{\mathbf{d}\in \Omega} a(\mathbf{d}) \mathrm{Pr}_{\mathrm{p}}(\mathbf{d}),
\end{aligned}
\end{equation}
where $a(\mathbf{x}^{(i)}, \mathbf{x}^{(j)}) \!=\! a(\mathbf{d})$ is the number of bit errors that occur when the vector $\mathbf{x}^{(i)}$ is transmitted and the vector $\mathbf{x}^{(j)}$ is detected, and $\Omega \!\triangleq\!\{\mathbf{d}\!=\!\mathbf{x}^{(i)}\!-\!\mathbf{x}^{(j)}\! \mid \mathbf{x}^{(i)}, \mathbf{x}^{(j)} \!\in\! \mathcal{X},\mathbf{x}^{(i)}\!\neq\! \mathbf{x}^{(j)} \} $ is the space of all possible distances.
$\mathrm{Pr}_{\mathrm{e}}^{\mathrm{PML}}$ and $\mathrm{Pr}_{\mathrm{e}}^{\mathrm{ML}}$ are thus expressed as
\begin{align}
    \mathrm{Pr}_{\mathrm{e}}^{\mathrm{PML}} & \le \frac{1}{|\mathcal{X}|} \sum_{\mathbf{d}\in \Omega} a(\mathbf{d}) \mathbb{E}\left[
    \frac{e^{-\frac{\norm{\dot{\mathbf{R}}\mathbf{d}}_{2}^{2}}{8\sigma^2 Q_{\mathrm{r}}}}}{12}\right] + a(\mathbf{d}) \mathbb{E}\left[\frac{e^{-\frac{\norm{\dot{\mathbf{R}}\mathbf{d}}_{2}^{2}}{12\sigma^2 Q_{\mathrm{r}}}}}{4}\right], 
    \label{eq:PML_empirical} \\
    \mathrm{Pr}_{\mathrm{e}}^{\mathrm{ML}} & \le \frac{1}{|\mathcal{X}|} \sum_{\forall \mathbf{d}} a(\mathbf{d}) \mathbb{E}\left[
    \frac{e^{-\frac{\norm{{\mathbf{R}}\mathbf{d}}_{2}^{2}}{8\sigma^2}}}{12}\right] + a(\mathbf{d}) \mathbb{E}\left[\frac{e^{-\frac{\norm{{\mathbf{R}}\mathbf{d}}_{2}^{2}}{12\sigma^2 }}}{4}\right].
    \label{eq:ML_empirical}
\end{align}


\section{Complexity analysis of studied detectors}
\label{sec:complexity}

Detector performance is crucial, but complexity and architectural constraints are equally important in \gls{thz} \gls{um-mimo} systems. Detector complexity affects computational resources, efficiency, latency, and scalability. Depending on application needs and hardware limitations, a detection scheme with lower complexity or better parallelization may be preferred. \gls{ml}, with its exhaustive lattice search, has exponential complexity, $\abs{\mathcal{X}}^{Q_{\mathrm{t}}}$, making it impractical for \gls{um-mimo} and higher modulation orders. Linear detectors like \gls{zf} reduce complexity to $\mathcal{O}\left(Q_{\mathrm{t}}^3 + Q_{\mathrm{t}}^2 Q_{\mathrm{r}}\right)$ (bounded by matrix inversion) but suffer performance loss with ill-conditioned channel matrices or high spatial correlation, which are common in \gls{thz} channels. However, such detectors perform well when adaptive array designs enforce channel orthogonality~\cite{Sarieddeen8765243}. \gls{sic} and its variants improve on linear detectors by sequentially detecting and canceling symbols, but are prone to error propagation.

Between exhaustive search and linear detectors, methods like \gls{sd} operate on a lattice subspace, with complexity defined by the search radius, which determines the nodes visited. Techniques such as pruning reduce the search tree size. K-Best \gls{sd}, with a fixed complexity of $2^{|\mathcal{X}|} + (Q_{\mathrm{t}} - 1) \cdot \text{K}^{\text{Best}} \cdot 2^{|\mathcal{X}|}$ determined by the number of active search paths ($K^{\mathrm{Best}}$), is advantageous for fixed latency hardware implementations. \gls{lr}-aided detection mitigates channel ill-conditioning by constructing a more orthogonal lattice. In \gls{lr}-aided \gls{zf} detection, while no universal upper bound exists for the number of \gls{lll} iterations in ill-conditioned \gls{mimo} channels \cite{4518202}, there exists an upper bound on its expected value $\mathbb{E}[\text{I}^{\text{LR}}]\!\! \leq\!\! 4 Q_{\mathrm{r}}^2\!\left(\log _2 \frac{Q_{\mathrm{r}}}{Q_{\mathrm{t}} - Q_{\mathrm{r}} + 1} \!+\!\frac{2.240}{\log _e t}\right)\!+\! 2 Q_{\mathrm{r}}$. Despite its non-deterministic complexity, \gls{lr} is particularly efficient in frequency-flat scenarios due to its one-time preprocessing overhead.

Parallelizable nonlinear detectors reduce system complexity and latency. Subspace detectors like \gls{lord}, with complexity $\mathcal{O}\left(Q_{\mathrm{r}}^2 |\mathcal{X}|\right)$, offer near-optimal performance with significantly lower complexity than \gls{ml}. However, tree-based architectures, including \gls{lord}, pose challenges for parallelization unless flattened and unrolled, increasing implementation complexity. Puncturing, especially in \gls{thz} channels, enhances parallelizability at a graceful performance cost, where the V-shape puncturing of $\mathbf{R}$ enables parallel processing and reduces computations \cite{Sarieddeen8186206}. The complexity of \gls{ssd} is given by $\mathcal{O}\left(Q_{\mathrm{r}}^2 |\mathcal{X}| - \Theta_{\mathrm{SSD}}\right)$, where $\Theta_{\mathrm{SSD}}$, the reduction factor compared to \gls{lord}, depends on the puncturing pattern and matrix dimensions \cite{Sarieddeen8186206}. This computation, performed once, can be stored in memory. With \gls{ssd}, the \gls{llr}s of specific symbols can be computed in parallel, enabling parallelization across the communication chain. In a $v \times v$ \gls{mimo} systems supporting parallel processing, \gls{ssd} reduces latency by a factor of $v$. Building on \gls{lord} and \gls{ssd}, the complexity of \gls{sqld} is expressed as $\mathcal{O}\left(Q_{\mathrm{r}}^2 |\mathcal{X}| - \frac{\eta}{Q_{\mathrm{t}}} \Theta_{\mathrm{SSD}}\right)$, where $\eta$ denotes the number of punctured layers. When $\eta = Q_{\mathrm{t}}$, \gls{sqld} reduces to \gls{ssd}, incurring only the additional overhead of layer sorting which results in performance enhancement.

In \gls{thz}-band communications, where large bandwidths contrast with limited baseband clock speeds, parallelism is key to bridging the “\gls{tbps} gap.” Subspace decomposition leverages the quasi-deterministic structure of \gls{thz} channels for scalable parallel processing, aligning \gls{tbps} data rate promises with hardware constraints while mitigating diversity losses \cite{sarieddeen2023bridging2}. \gls{ssd} is inherently scalable, managing larger system dimensions without a proportional increase in complexity. Its robustness arises from isolating errors within subspaces, reducing error propagation typical of sequential schemes like \gls{sic}. In \gls{thz} channels with high spatial correlation and limited scattering, \gls{ssd} effectively leverages channel characteristics. Its architecture aligns with \gls{thz} hardware constraints by mapping independent subspaces to separate cores, optimizing resource utilization, and minimizing latency. Techniques such as pipelining and fixed-point arithmetic further enable real-time operation at \gls{thz} symbol rates \cite{douillard2019next}.

\section{Wideband design and proposed extensions}
\label{sec:wideband_reuse}

\begin{table*}[ht]
    \centering
    \footnotesize
    \caption{FLOPs complexity comparison of \gls{sic}, \gls{lord}, and \gls{ssd} detectors with and without \gls{thz} wideband \gls{qrd} reuse strategy}
    \resizebox{\textwidth}{!}{%
    \begin{tabular}{|p{1.6cm}|c|c|p{4.5cm}|p{5.5cm}|p{6.5cm}|}
        \hline
          & \textbf{Scheme} &  & \textbf{SIC} & \textbf{LORD} & \textbf{SSD (WRD-only)} \\
        \hline
        \multirow{4}{*}{\textbf{No Reuse}} &
        \multirow{2}{*}{\textbf{RADD}} & \textbf{AR} &
        $M\!\left(\theta_{\mathrm{QRD}}^{\text{RADD}} + 2 Q_t^2 - 2 Q_t\right)$ &
        $M\!\left(Q_t\,\theta_{\mathrm{QRD}}^{\text{RADD}} + Q_t|\mathcal{M}|\,(4 Q_t^2 + 2 Q_t - 1)\right)$ &
        $M\!\Big(Q_t\,\theta_{\mathrm{WRD}}^{\text{RADD}} + Q_t|\mathcal{M}|\,(4 Q_t^2 + 2 Q_t - 1) - Q_t\!\left(|\mathcal{M}|(Q_t^2 - 3 Q_t + 2) - (4 Q_t^2 + 4 Q_t - 2)\right)\Big)$ \\
        \cline{3-6}
        & & \textbf{TI} &
        $\theta_{\mathrm{QRD}}^{\text{RADD}} + 2 Q_t^2 - 2 Q_t$ &
        $Q_t\,\theta_{\mathrm{QRD}}^{\text{RADD}} + Q_t|\mathcal{M}|\,(4 Q_t^2 + 2 Q_t - 1)$ &
        $Q_t\,\theta_{\mathrm{WRD}}^{\text{RADD}} + Q_t|\mathcal{M}|\,(4 Q_t^2 + 2 Q_t - 1) - Q_t\!\left(|\mathcal{M}|(Q_t^2 - 3 Q_t + 2) - (4 Q_t^2 + 4 Q_t - 2)\right)$ \\
        \cline{2-6}
        &
        \multirow{2}{*}{\textbf{RMUL}} & \textbf{AR} &
        $M\!\left(\theta_{\mathrm{QRD}}^{\text{RMUL}} + 2 Q_t^2\right)$ &
        $M\!\left(Q_t\,\theta_{\mathrm{QRD}}^{\text{RMUL}} + Q_t|\mathcal{M}|\,(4 Q_t^2 + 4 Q_t)\right)$ &
        $M\!\Big(Q_t\,\theta_{\mathrm{WRD}}^{\text{RMUL}} + Q_t|\mathcal{M}|\,(4 Q_t^2 + 4 Q_t) - Q_t\!\left(|\mathcal{M}|(2 Q_t^2 - 6 Q_t + 4) - (4 Q_t^2 + 4 Q_t)\right)\Big)$ \\
        \cline{3-6}
        & & \textbf{TI} &
        $\theta_{\mathrm{QRD}}^{\text{RMUL}} + 2 Q_t^2$ &
        $Q_t\,\theta_{\mathrm{QRD}}^{\text{RMUL}} + Q_t|\mathcal{M}|\,(4 Q_t^2 + 4 Q_t)$ &
        $Q_t\,\theta_{\mathrm{WRD}}^{\text{RMUL}} + Q_t|\mathcal{M}|\,(4 Q_t^2 + 4 Q_t) - Q_t\!\left(|\mathcal{M}|(2 Q_t^2 - 6 Q_t + 4) - (4 Q_t^2 + 4 Q_t)\right)$ \\
        \hline
        \multirow{4}{*}{\textbf{With Reuse}} &
        \multirow{2}{*}{\textbf{RADD}} & \textbf{AR} &
        $M\!\left((1{-}P)\,\theta_{\mathrm{QRD}}^{\text{RADD}} + 2 Q_t^2 - 2 Q_t\right)$ &
        $M\!\left(Q_t(1{-}P)\,\theta_{\mathrm{QRD}}^{\text{RADD}} + Q_t|\mathcal{M}|\,(4 Q_t^2 + 2 Q_t - 1)\right)$ &
        $M\!\Big(Q_t(1{-}P)\,\theta_{\mathrm{WRD}}^{\text{RADD}} + Q_t|\mathcal{M}|\,(4 Q_t^2 + 2 Q_t - 1) - Q_t\!\left(|\mathcal{M}|(Q_t^2 - 3 Q_t + 2) - (4 Q_t^2 + 4 Q_t - 2)\right)\Big)$ \\
        \cline{3-6}
        & & \textbf{TI} &
        $(1{-}P)\,\theta_{\mathrm{QRD}}^{\text{RADD}} + 2 Q_t^2 - 2 Q_t$ &
        $Q_t(1{-}P)\,\theta_{\mathrm{QRD}}^{\text{RADD}} + Q_t|\mathcal{M}|\,(4 Q_t^2 + 2 Q_t - 1)$ &
        $Q_t(1{-}P)\,\theta_{\mathrm{WRD}}^{\text{RADD}} + Q_t|\mathcal{M}|\,(4 Q_t^2 + 2 Q_t - 1) - Q_t\!\left(|\mathcal{M}|(Q_t^2 - 3 Q_t + 2) - (4 Q_t^2 + 4 Q_t - 2)\right)$ \\
        \cline{2-6}
        &
        \multirow{2}{*}{\textbf{RMUL}} & \textbf{AR} &
        $M\!\left((1{-}P)\,\theta_{\mathrm{QRD}}^{\text{RMUL}} + 2 Q_t^2\right)$ &
        $M\!\left(Q_t(1{-}P)\,\theta_{\mathrm{QRD}}^{\text{RMUL}} + Q_t|\mathcal{M}|\,(4 Q_t^2 + 4 Q_t)\right)$ &
        $M\!\Big(Q_t(1{-}P)\,\theta_{\mathrm{WRD}}^{\text{RMUL}} + Q_t|\mathcal{M}|\,(4 Q_t^2 + 4 Q_t) - Q_t\!\left(|\mathcal{M}|(2 Q_t^2 - 6 Q_t + 4) - (4 Q_t^2 + 4 Q_t)\right)\Big)$ \\
        \cline{3-6}
        & & \textbf{TI} &
        $(1{-}P)\,\theta_{\mathrm{QRD}}^{\text{RMUL}} + 2 Q_t^2$ &
        $Q_t(1{-}P)\,\theta_{\mathrm{QRD}}^{\text{RMUL}} + Q_t|\mathcal{M}|\,(4 Q_t^2 + 4 Q_t)$ &
        $Q_t(1{-}P)\,\theta_{\mathrm{WRD}}^{\text{RMUL}} + Q_t|\mathcal{M}|\,(4 Q_t^2 + 4 Q_t) - Q_t\!\left(|\mathcal{M}|(2 Q_t^2 - 6 Q_t + 4) - (4 Q_t^2 + 4 Q_t)\right)$ \\
        \hline
    \end{tabular}}%
    \label{tab:flops_comparison_combined}
\end{table*}

In \gls{um-mimo} wideband \gls{thz} systems, detection is typically performed per subcarrier. Building on the complexity analysis in Sec.\ref{sec:complexity}, which quantifies the \gls{qrd} cost in the studied detectors, for \gls{qrd}-based detectors, including subspace and \gls{sic}-based schemes, performing \gls{qrd} for each subcarrier introduces significant computational overhead, which worsens as the number of subcarriers increases. This overhead is even more severe for detectors like \gls{lord} and \gls{ssd}, where \gls{qrd} is required for every layer permutation. These costs are especially critical at \gls{thz}, where wide bandwidths and large arrays impose stringent computational and power constraints. To mitigate this burden while maintaining performance, we propose a channel-matrix reuse strategy across multiple subcarriers. Specifically, we reuse a single \gls{qrd} computed on a representative subset of subcarriers within a coherence bandwidth and propagate it to the remaining subcarriers, thereby reducing \gls{flops}, latency, and energy with minimal performance degradation.

Our approach partitions the system subcarriers into blocks, where detection preprocessing operations are computed once and reused for subsequent subcarriers within each block. This reuse process leverages results from the immediately preceding subcarrier. We define two key parameters: $\Delta m$ represents the number of subcarriers over which an operation, such as \gls{qrd}, is reused before recomputation, and $P \in [0,1]$ denotes the fractional reduction, indicating the proportion of subcarriers where \gls{qrd} is reused. For example, $P = 0.25$ corresponds to a $25\%$ reduction, meaning that for every four subcarriers, the \gls{qrd} is computed three times and reused once. The relationship between the total number of subcarriers, $M$, $\Delta m$, and $P$ is
\begin{equation}
    \Delta m = \frac{M}{\text{Numb. of recomputed subcarriers}} = \frac{1}{1- P}.
\end{equation}
For example, for a 20\% reduction in \gls{qrd} computations with $M = 10$ subcarriers, we compute \gls{qrd} for 80\% (i.e., 8 subcarriers) and reuse it for the remaining 20\% (i.e., 2 subcarriers). Thus, the reuse factor is $\Delta m = \frac{10}{8} = 1.25$. In static \gls{los} THz scenarios, quasi-deterministic channels result in a very high $\Delta m$. In other scenarios, $\Delta m$ dictates a trade-off between computational complexity, power consumption, and detection accuracy. This strategy is especially suited to high-throughput scenarios where real-time processing requirements necessitate avoiding redundant computations.

The computational complexity of the \gls{sic}, \gls{lord}, and \gls{ssd} detectors is first analyzed without acounting for reduction in \gls{qrd} decomposition operations in Table \ref{tab:flops_comparison_combined}. We differentiate between the timing (TI) complexity and arithmetic (AR) complexity, as each captures a distinct aspect of computational performance. This distinction is essential for understanding parallelizability: algorithms with high AR but low TI can benefit from parallel execution, whereas high TI imposes inherent latency limits. In this baseline scenario with $\Delta m = 1$, each subcarrier undergoes full \gls{qrd} decomposition, yielding the highest \gls{flops} count. Assumptions include mapping complex operations to their real equivalents: complex multiplication is $4$ \gls{rmul} and $2$ \gls{radd}, and complex addition is $2$ \gls{radd}. The \gls{flops} for \gls{qrd} are given by: $\theta_{\mathrm{QRD}}^{\text{RADD}} = 4Q_{\mathrm{r}} Q_{\mathrm{t}}^2 - Q_{\mathrm{t}}^2 - Q_{\mathrm{t}}$ and $\theta_{\mathrm{QRD}}^{\text{RMUL}} = 4Q_{\mathrm{r}} Q_{\mathrm{t}}^2 + 3Q_{\mathrm{t}}^2$. Similarly, for WRD: $\theta_{\mathrm{WRD}}^{\text{RADD}} = \frac{16}{3}Q_{\mathrm{r}} Q_{\mathrm{t}}^3 - 10Q_{\mathrm{r}} Q_{\mathrm{t}}^2 + \frac{8}{3}Q_{\mathrm{r}} Q_{\mathrm{t}} - 8Q_{\mathrm{r}}$ and $\theta_{\mathrm{WRD}}^{\text{RMUL}} = \frac{16}{3}Q_{\mathrm{r}} Q_{\mathrm{t}}^3 - 7Q_{\mathrm{r}} Q_{\mathrm{t}}^2 + \frac{8}{3}Q_{\mathrm{r}} Q_{\mathrm{t}} - 20Q_{\mathrm{r}}$. The FLOPs savings from WRD are expressed as $\theta_1^{\text{RADD}} = Q_{\mathrm{t}}^2 - 3Q_{\mathrm{t}} + 2$ and $\theta_1^{\text{RMUL}} = 2Q_{\mathrm{t}}^2 - 6Q_{\mathrm{t}} + 4$ \cite{Sarieddeen8186206}. These expressions quantify the computational overhead from the \gls{qrd} and WRD processes.

Table \ref{tab:flops_comparison_combined} extends this analysis by introducing the reduction percentage $P$, quantifying the reuse of \gls{qrd}s across subcarriers. Furthermore, Table \ref{tab:flops_comparison_combined} presents each detector's modified \gls{flops} expressions with a $P$ percent reduction. As $P$ increases, \gls{radd} and \gls{rmul} counts decrease proportionally. The \gls{sic} detector benefits from \gls{qrd} reuse, with its \gls{flops} scaling linearly by a factor of $(1 - P)$. In contrast, \gls{lord} experiences multiple reductions due to its multi-layered architecture, where each layer's \gls{flops} scale by $(1 - P)$, resulting in a more significant overall reduction compared to \gls{sic}. The reduction applied to \gls{qrd} similarly extends to the \gls{wrd}-based \gls{ssd}; however, since \gls{ssd} already requires fewer computations, the resulting complexity gains are even more pronounced, making it the most favorable detector under the proposed strategy.

\begin{figure}[ht!]
  \centering
    \includegraphics[width=0.48\textwidth]{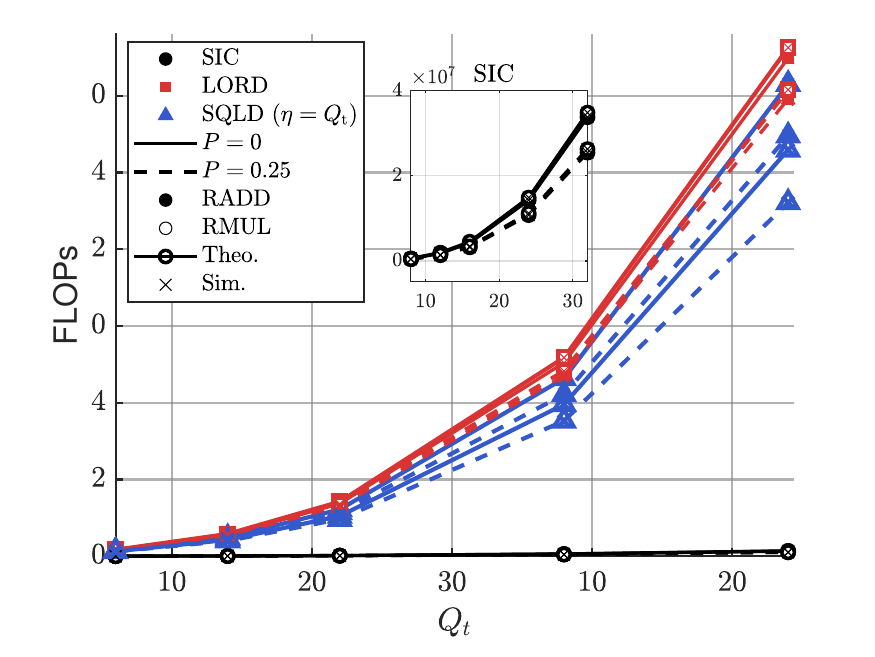}
 \caption{FLOPs vs $Q_t$, $|\mathcal M|{=}64$, $M{=}256$)}
  \label{fig:flops_plot}
\end{figure}
Fig.~\ref{fig:flops_plot} illustrates the \gls{flops} scaling of the three detectors. \gls{sic} exhibits the expected $\mathcal{O}(Q_t^2)$ growth, as confirmed by the zoomed inset, and therefore remains far below the layered schemes across all $Q_{\mathrm{t}}$.
Both \gls{lord} and \gls{sqld} scale as $\mathcal{O}(|\mathcal{M}|Q_t^3)$, reflecting the \gls{qrd}, \gls{wrd}, and distance metric computations. The curves for $P=0$ (solid) and $P=0.25$ (dashed) show that only the factorization cost is reduced by reuse, while the candidate loops are unaffected. Here SQLD with $\eta = Q_{\mathrm{t}}$ corresponds to SSD. Importantly, SQLD modifies LORD’s complexity in two ways: it replaces the QRD costs ($Q_t\theta_{\mathrm{QRD}}^{\text{RMUL}}$ and $Q_t\theta_{\mathrm{QRD}}^{\text{RADD}}$) by the lighter WRD costs ($Q_t\theta_{\mathrm{WRD}}^{\text{RMUL}}$ and $Q_t\theta_{\mathrm{WRD}}^{\text{RADD}}$), and it eliminates $Q_t|\mathcal{M}|\theta_{1}$ operations in the per–candidate loop, where $\theta_{1}$ represents the puncturing savings. These changes lower the cubic coefficient and, together with WRD’s structured $R$ matrix that shortens data dependencies, make SQLD more parallelizable. With reuse ($P=0.25$), the relative advantage of SQLD widens: the dashed blue curves fall consistently below the dashed red curves. Finally, the excellent agreement between theoretical FLOP counts and simulation results ($\times$ markers) validates the analysis. In summary, SQLD achieves lower FLOP complexity (via a controlled $\eta$, $\theta_{1}$ and avoiding $\theta_{2}$), inducing higher parallelizability.

\section{Simulation Results}
\label{sec:sim_res_disc}

This section evaluates the performance metrics obtained using Monte Carlo simulations, across the channel models detailed in Sec.~\ref{subsec:thz_channel_models}. All channels are generated with the TeraMIMO simulator~\cite{tarboush9591285} where the simulation parameters are configured based on validated measurement campaigns for both THz indoor and outdoor scenarios~\cite{tarboush9591285,papasotiriou2021experimentally,papasotiriou2023outdoor}. The receiver \gls{snr} is defined per spatial stream and per symbol at the detector input (after \gls{rf} combining and before soft-output detection). With total transmit power \(P_{\mathrm{Tx,Tot}}\) equally allocated across the \(N_{\mathrm s}\) spatial streams and large-scale loss factor \(\mathrm{PL}\) (free-space plus molecular absorption), the per-stream symbol \gls{snr} is defined as
\begin{equation}
\label{eq:snr_stream}
\mathrm{SNR} \;\triangleq\; \frac{P_{\mathrm{Tx,Tot}}\times \mathrm{PL}}{N_{\mathrm s}\,\sigma^{2}},
\end{equation}
where \(\sigma^{2}\) is the noise power integrated over the system bandwidth. This choice is consistent with the detection model in \eqref{eq:sys_model_secondline}.  Since the focus of this work is on soft-output detection, channel coding is incorporated into the simulation framework to provide a realistic performance evaluation. In particular, Turbo coding is applied with a code length of 256 bits and an encoding rate of 1/3, ensuring robust error correction under varying \gls{snr} conditions.

\subsection{THz MIMO detection}

\begin{figure*}[t!]%
 \centering
 \subfloat[16x16 MIMO, QPSK, \gls{thz} indoor channel \\ $F_{\mathrm{c}}=0.142$ THz, and $D=3.15$ m ]{\includegraphics[width=0.49\linewidth]{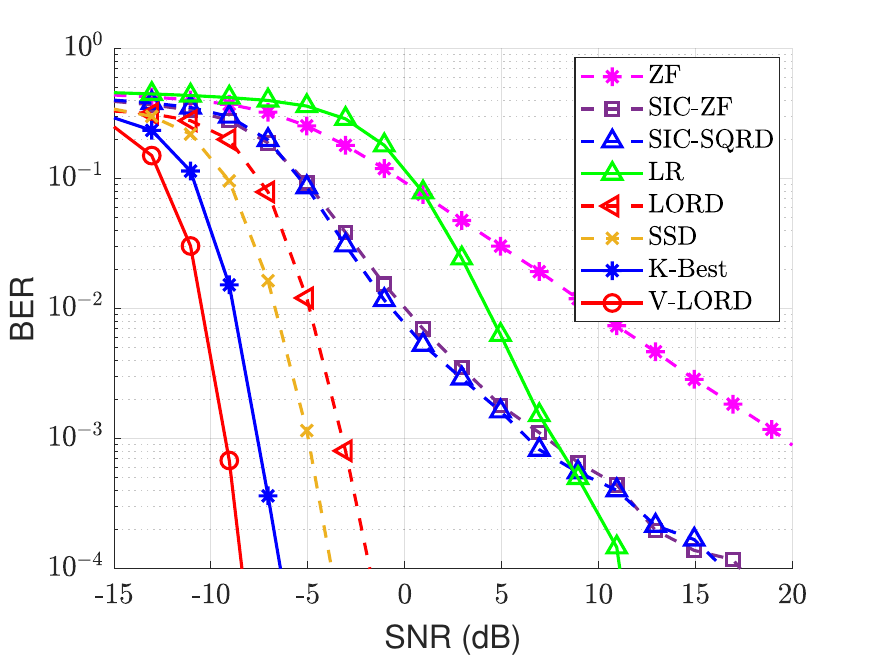}\label{fig:indoor}}
 \subfloat[16x16 MIMO, QPSK, \gls{thz} outdoor channel \\ $F_{\mathrm{c}}=0.142$ THz, $D=64$ m, and $K=3$]{\includegraphics[width=0.49\linewidth]{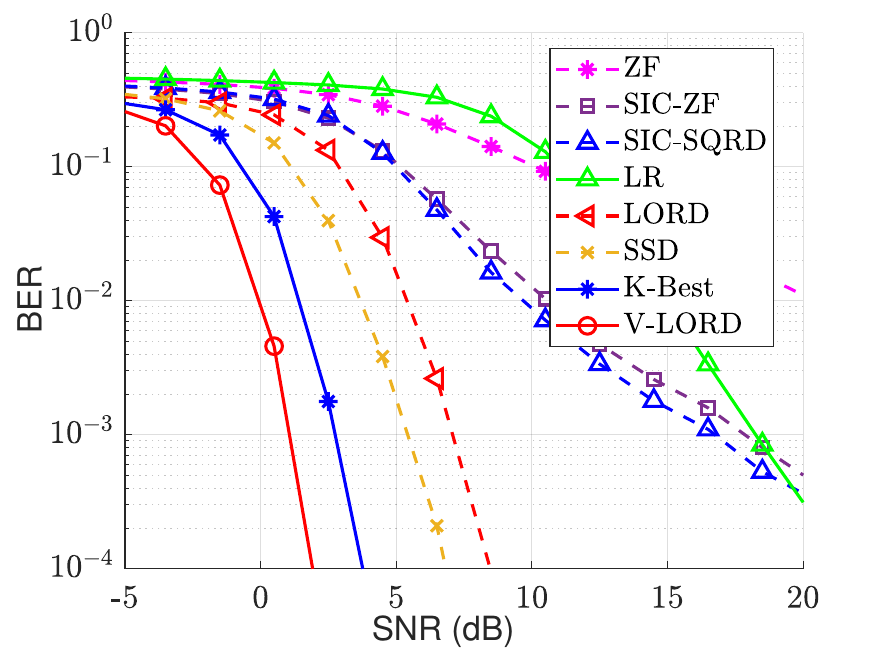}\label{fig:outdoor}}\\
 \subfloat[16x16 \gls{um-mimo}, 25x25 \gls{ae}s, 16-QAM, \gls{thz} data center \\ channel, $F_{\mathrm{c}}=0.3$ THz, and $D=1.75$ m]{\includegraphics[width=0.49\linewidth]{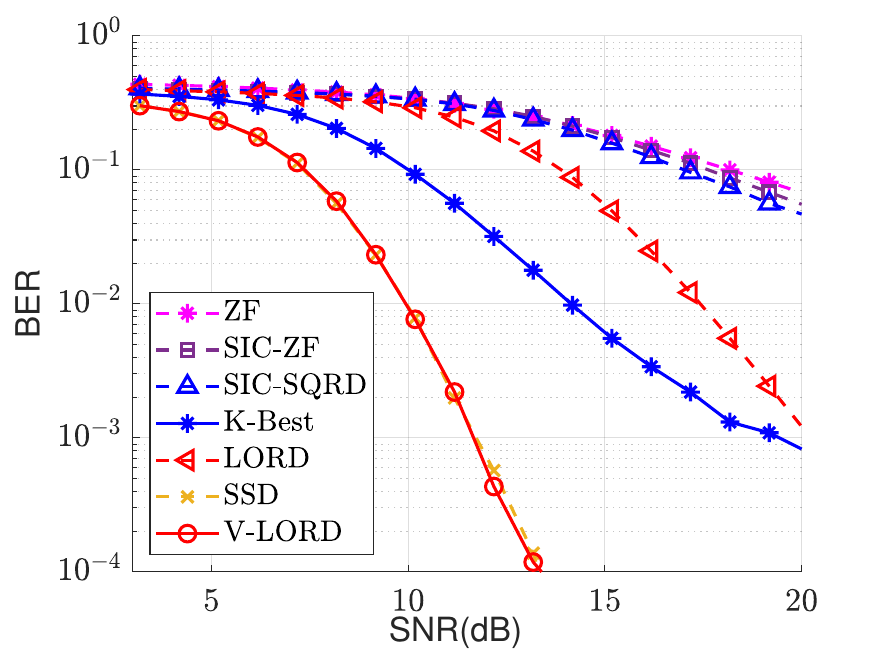}\label{fig:datacenter}}%
 \subfloat[16x16 \gls{um-mimo}, 25x25 \gls{ae}s, 16-QAM, \gls{thz} \gls{los} channel, \\ $F_{\mathrm{c}}=0.45$ THz, and $D=1$ m]{\includegraphics[width=0.49\linewidth]{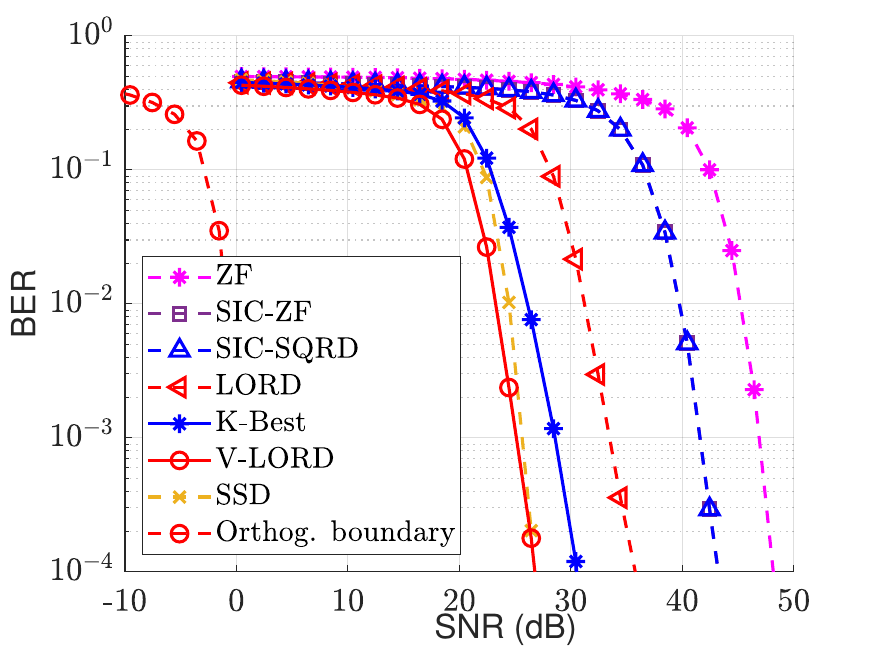}\label{fig:los_orthog_non_orthogonal}}
 \caption{Performance comparison of studied detectors under different \gls{thz} channel models.}%
 \label{detectors_channels}%
\end{figure*}

Simulation results are provided for both \gls{nf} and \gls{ff} \gls{thz} communication channels.  
In the \gls{nf} regime, we adopt the \gls{hspwm} for the \gls{aosas} and detector performance is evaluated under two conditions: a data center channel at \unit[0.3]{\gls{thz}} \cite{tarboush9591285} and a \gls{los} channel at \unit[0.45]{\gls{thz}}, both with transmission distances shorter than \unit[2]{m}.  
For the \gls{ff} regime, we use \gls{pwm} and the performance is investigated for indoor and outdoor channel conditions at \unit[0.142]{\gls{thz}} \cite{papasotiriou2021experimentally,papasotiriou2023outdoor}.

\subsubsection{Far-field THz MIMO channels}

We first illustrate results for indoor and outdoor channel models in Fig.~\ref{fig:indoor} and Fig.~\ref{fig:outdoor}, respectively. The performance order remains consistent across both scenarios, with key differences in the \gls{snr} region and the gap between \gls{ssd} and \gls{v-lord}. Although both scenarios operate at \unit[0.142]{\gls{thz}}, the communication distance varies: \unit[3.15]{m} indoors and \unit[64]{m} outdoors. This difference, coupled with the absence of beamforming, accounts for the \unit[30]{dB} performance gap between the two. While \gls{zf} is effective at high \gls{snr}, it amplifies noise in poorly conditioned channels, which is common in \gls{thz} environments. \gls{sic} mitigates interference more effectively by sequentially decoding and canceling detected symbols. LR enhances detection by transforming the channel into a more orthogonal lattice structure. LR-aided \gls{zf} detection improves performance by \unit[10–15]{dB} at \unit[$10^{-3}$]{BER}. SSD, leveraging puncturing, outperforms LORD by approximately \unit[3]{dB} indoors and \unit[2]{dB} outdoors at $10^{-4}$ \gls{ber}. With higher channel correlation, SSD gains further advantage as it breaks channel-layer dependencies through puncturing, a trend observed across all \gls{thz} channel models in this work.

\subsubsection{Near-field THz MIMO channels} 

Fig.~\ref{fig:datacenter} shows the performance of the evaluated detectors in a \gls{thz} data center channel model with a 16x16 \gls{sa} and 25x25 \gls{ae}s \gls{um-mimo} system. The \gls{zf} detector performs the worst, exhibiting a performance gap of approximately \unit[20]{dB} at a \gls{ber} of $10^{-2}$, which increases at lower \gls{ber} levels compared to subspace detectors. Among low-complexity schemes, SIC-\gls{zf} and SIC-\gls{sqrd} offer only limited improvements over \gls{zf}. Specifically, SIC-\gls{sqrd} provides an additional gain of \unit[0.5]{dB} at a \gls{ber} of $10^{-2}$ with a modest increase in complexity. Among the subspace detectors, SSD shows a performance comparable to that of \gls{v-lord}, highlighting its competitiveness in \gls{thz} environments due to its significantly lower complexity. Subspace decomposition techniques, such as channel puncturing, further mitigate the impact of high channel correlation in the \gls{thz} regime.

Next, we examine the \gls{los} channels in Fig.~\ref{fig:los_orthog_non_orthogonal}. \gls{thz} \gls{um-mimo} designs are inherently reconfigurable, and we distinguish between orthogonal and non-orthogonal \gls{los} based on whether spatial tuning of $\Delta$ is supported. Near-orthogonality can be achieved under specific conditions at an optimal \gls{sa} separation, $\Delta_{\mathrm{opt}}\!=\!(zDc)/(Q_{\mathrm{t}}F_{\mathrm{c}})$, for odd $z$ values \cite{Sarieddeen8765243}, where $c$ is the speed of light, $D$ is the communication distance, and $F_{\mathrm{c}}$ is the operating frequency. Two key observations emerge from these simulations. First, SSD narrows the performance gap with V-LORD at a \gls{ber} of $10^{-4}$ and below, achieving an improvement of approximately \unit[10]{dB} over LORD. This demonstrates SSD's ability to mitigate correlation through puncturing, making it a promising candidate for subspace detection in \gls{thz} ill-conditioned \gls{mimo} and \gls{um-mimo} channels due to its complexity and performance advantages. Second, \gls{zf} faces difficulties in decoupling signals in highly ill-conditioned channels, limiting its effectiveness. However, in orthogonal \gls{los} scenarios enabled through spatial tuning, all detection schemes perform similarly, as shown by the red dotted curve on the left of Fig.~\ref{fig:los_orthog_non_orthogonal}. Under near-orthogonal or fully orthogonal conditions, \gls{zf} shows marked improvement, closing the gap with sub-optimal detectors and becoming a viable detection choice.

\subsection{ Performance of proposed SQLD}
\begin{figure}[t!]
  \centering
    \includegraphics[width=0.48\textwidth]{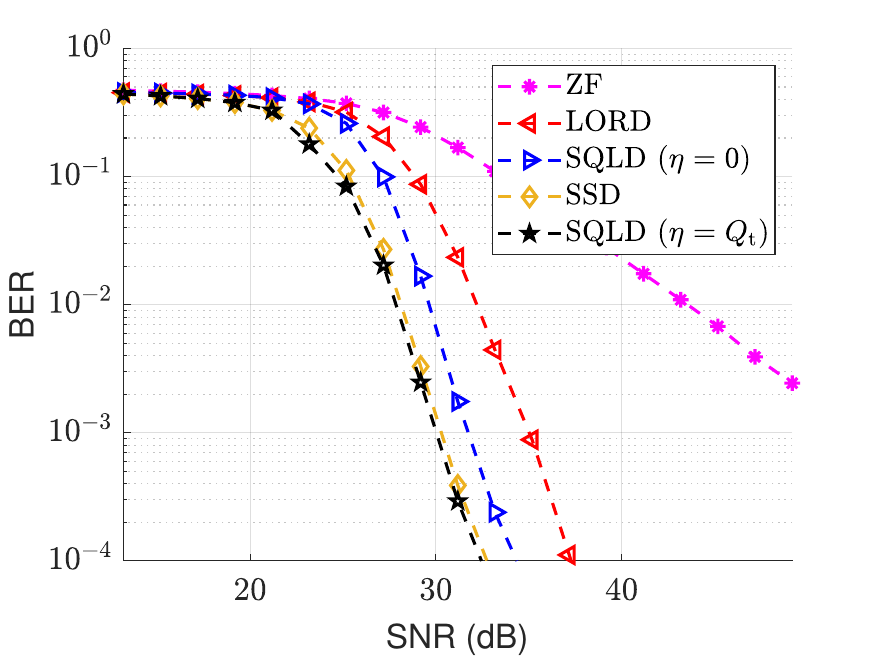}
 \caption{Performance comparison between LORD, SSD, and the proposed SQLD: 16x16 MIMO, 4-QAM, \gls{thz} data center channel, $F_{\mathrm{c}}=0.142$ THz, and $D=1.75$ m.}
  \label{fig:SSD_hard}
\end{figure}

Fig. \ref{fig:SSD_hard} compares the performance of \gls{lord} and \gls{ssd} with the proposed SQLD for $\eta = 0$ and $\eta = Q_{\mathrm{t}}$ under the \gls{thz} data center channel. The simulation uses a 16x16 \gls{mimo} system with 4-\gls{qam} modulation and hard-output detectors to evaluate the lowest complexity configuration of Algorithm \ref{alg:SQLD}.

Two key insights emerge from the results. First, SQLD with $\eta = 0$ improves \gls{ber} by approximately 4 dB over \gls{lord} at a \gls{ber} of 
 \(10^{-4}\) in the ill-conditioned \gls{thz} channel. This highlights the benefits of sorting with \gls{lord} in high-dimensional \gls{um-mimo} systems, where channel conditioning is challenging. The smaller improvement of SQLD over \gls{lord} in the Rayleigh channel \cite{Sarieddeen8186206} further suggests that SQLD is particularly effective in high-frequency \gls{thz} environments. Second, SQLD with full parallelizability $\eta = Q_{\mathrm{t}}$ achieves a \unit[0.5]{dB} gain. While still an improvement, its limited effect likely stems from suboptimal sorting in the punctured \gls{ssd} channel or the degradation caused by puncturing.

\subsection{THz LoS wideband UM-MIMO}
\begin{figure}[t!]
  \centering
    \includegraphics[width=0.48\textwidth]{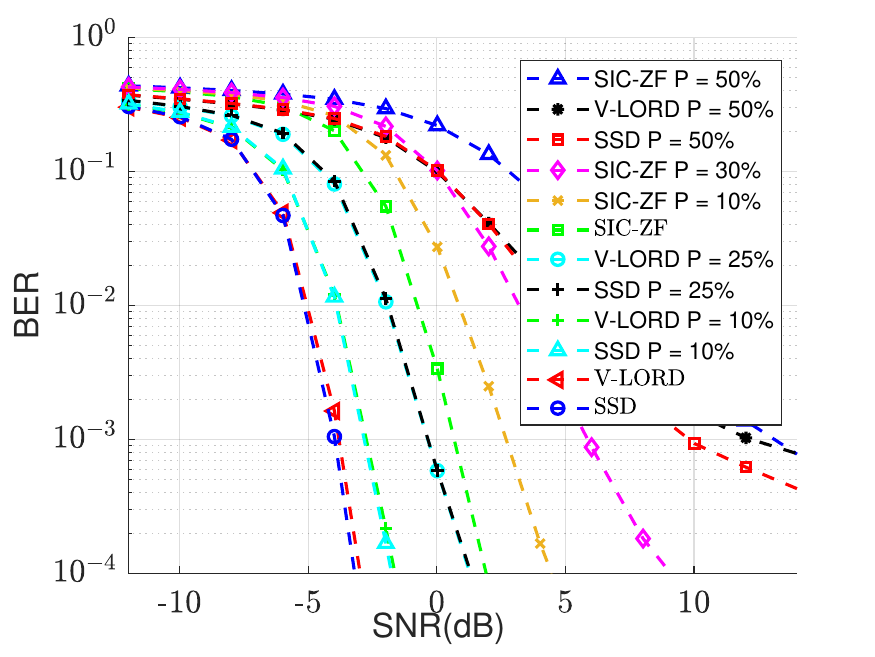}
 \caption{Performance comparison in wideband \gls{los} communications: 4x4 \gls{um-mimo} with 256x16 \gls{ae}s, $F_{\mathrm{c}}=0.142$ THz, and $D=1$ m.}
  \label{fig:performance_analysis_wideband}
\end{figure}

Fig.~\ref{fig:performance_analysis_wideband} examines the reuse strategy from Sec.~\ref{sec:wideband_reuse} in a \gls{thz} wideband \gls{um-mimo} setup: a $4 \times 4$ \gls{um-mimo} system with 256 \gls{ae}s at the transmitter and 16 \gls{ae}s at the receiver, operating in \gls{los} over a 1 m link at a central frequency of 0.325 \gls{thz} and 32 GHz bandwidth. Using TeraMIMO, we simulate the \gls{thz} channel with the beam-split effect for wideband modeling. The reuse strategy reduces \gls{qrd} computations, offering substantial computational savings. The \gls{ber} curves show that performance degradation remains minimal with higher reuse (e.g., $P=50\%$) at moderate \gls{snr}, but becomes noticeable at higher \gls{snr}s.

The impact of \gls{qrd} reuse in \gls{v-lord} is promising, with a minor \gls{ber} gap between \gls{v-lord} with $P=25\%$ and the baseline, indicating minimal performance loss with reuse. However, higher reuse percentages like $P=50\%$ lead to more noticeable \gls{ber} degradation at higher \gls{snr}s, suggesting an upper limit for reuse without compromising reliability in high-\gls{snr} regimes. A similar trend is observed for \gls{ssd}, where the fine-grained accuracy of \gls{qrd} is more critical at high \gls{snr}. The $P=25\%$ reduction strikes an optimal balance, maintaining low \gls{ber} while effectively reducing computational demands. This analysis shows that, with carefully selected reuse levels, wideband \gls{thz} systems can achieve significant computational savings with manageable \gls{ber} trade-offs, making them feasible for practical high-data-rate applications at \gls{thz} frequencies. Note that under extreme flat fading conditions, the proposed wideband reuse strategy maximizes complexity reduction while incurring minimal performance degradation, as the channel's frequency response remains nearly constant over a wide bandwidth. Notably, a single \gls{qrd} can be reused across all subcarriers, significantly reducing computational overhead. This is particularly beneficial for high-data-rate \gls{thz} systems, where power efficiency and low latency are critical.

\subsection{MIMO Data detection under beam split}
\begin{figure}[t!]
  \centering
    \includegraphics[width=0.48\textwidth]{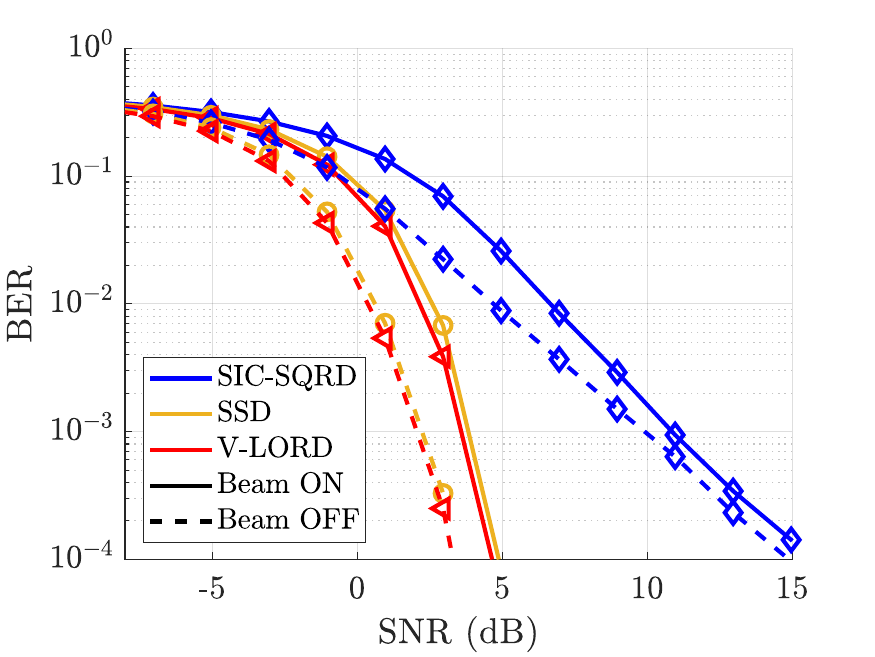}
 \caption{Performance comparison with and without beam split effect: 4x4 SAs 256x16 \gls{um-mimo}, \gls{thz} \gls{los} and \gls{los}+MP channels, $F_{\mathrm{c}}=0.325$ THz, and $D=1$ m.}
  \label{fig:beam_split}
\end{figure}

In wideband \gls{thz} systems, beam split (or beam squint) arises from the frequency-independent nature of analog RF phase shifters \cite{tarboush9591285}. Since the analog precoder and combiner are optimized at the carrier frequency~$F_{\mathrm{c}}$, the same phase settings become increasingly misaligned at subcarriers~$f_m$. The spatial frequency scales approximately as~$f_m/F_{\mathrm{c}}$, such that a beam designed to steer toward~$\theta_c$ at~$F_{\mathrm{c}}$ effectively points toward~$\theta_m$, where $\sin\theta_m \approx (f_m/F_{\mathrm{c}})\sin\theta_c$. This frequency-dependent steering causes the mainlobe to deviate from its intended direction, reducing coherent array gain and introducing inter-stream interference across subcarriers. The severity of this effect grows with both the array aperture and the fractional bandwidth~$B/F_{\mathrm{c}}$. Fig.~\ref{fig:beam_split} compares detection with Beam split enabled (solid) versus disabled (dashed). The configuration simulated is a \gls{thz} \gls{los} 4x4 \gls{sa}s and 256x16 \gls{ae}s \gls{um-mimo} system operating at \unit[0.325]{THz} and with a bandwidth of \unit[0.0325]{THz}. Across the entire SNR range, the solid curves lie above the dashed ones, indicating a consistent performance loss due to beam split. 
The penalty is detector dependent: SIC-SQRD is most sensitive (roughly a \unit[1--2]{dB} \gls{snr} loss at \gls{ber} between \(10^{-3}\) and \(10^{-4}\)), whereas LORD and SSD incur a smaller shift (about $0.3$–$0.6$\,dB in the same BER region). This behavior aligns with the underlying physics: frequency-dependent steering reduces coherent array gain, which worsens the effective channel conditioning due to reduced spatial orthogonality. SIC relies heavily on orthogonality for reliable cancellation and therefore degrades more, while near-ML schemes (LORD/SSD) are more robust and preserve their steeper high-SNR slopes. Overall, in this configuration beam split acts as a non-negligible impairment, with the largest impact on SIC-SQRD and a noticeably smaller (but still visible) impact on LORD/SSD.

\subsection{MIMO detection under hybrid beamforming}
\label{subsec:hybrid_sim}
\begin{figure}[t!]
  \centering
    \includegraphics[width=0.48\textwidth]{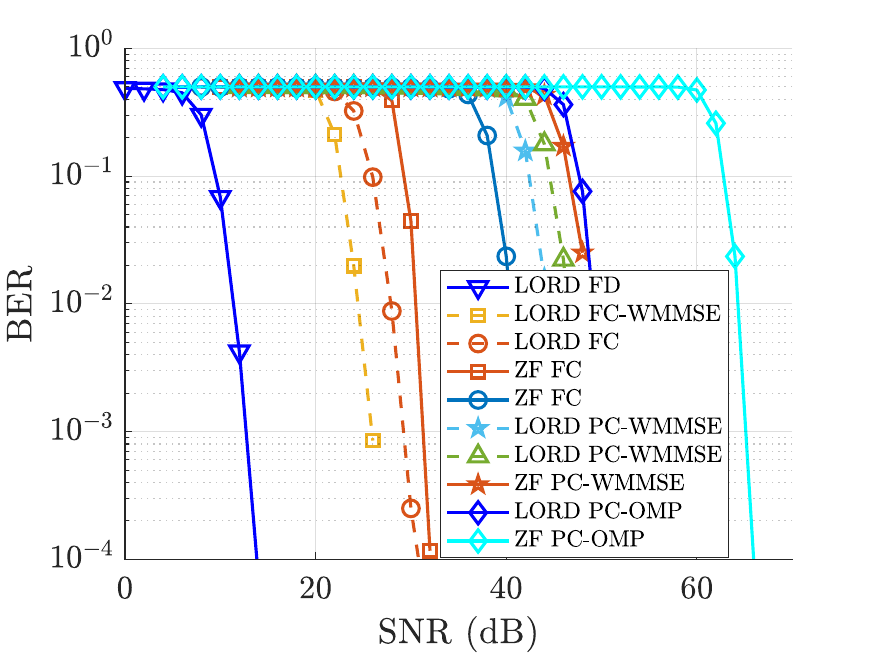}
 \caption{Performance comparison under hybrid beamforming: 16x16 \gls{sa}s 16x16 \gls{um-mimo}, \gls{thz} \gls{los} channel, $F_{\mathrm{c}}=0.3$ THz, and $D=1$ m.}
  \label{fig:precoding}
\end{figure}

In our work, detection is performed on the baseband channel $\mathbf{H}[m]$ defined in \eqref{eq:H_detect}, with the AE-domain channel $\widecheck{\mathbf{H}}[m]$ first mapped to the \gls{sa}-level effective channel $\widehat{\mathbf{H}}_{\mathrm{SA}}[m]$ via \eqref{eq:H_SA_def}. To evaluate the robustness of \gls{thz} \gls{mimo} detection under different precoding/combining designs, we compare five coherent chains—\gls{fd}, \gls{fc}, \gls{fc}+\gls{wmmse}, \gls{pc}+\gls{omp}, and \gls{pc}+\gls{wmmse}—and report the \gls{ber} for \gls{zf} and \gls{lord}
\begin{itemize}
    \item \gls{fd}: In this ideal bound, we set identity \gls{rf} precoding and combining matrices $\mathbf{F}_{\mathrm{RF}}=\mathbf{I}_{Q_{\mathrm t}Q^2}$ and $\mathbf{W}_{\mathrm{RF}}=\mathbf{I}_{Q_{\mathrm r}Q^2}$ so that $\widehat{\mathbf{H}}_{\mathrm{SA}}[m]=\widecheck{\mathbf{H}}[m]$. Then, we compute for each subcarrier the \gls{svd} $\widecheck{\mathbf{H}}[m]=\mathbf{U}_m\boldsymbol{\Sigma}_m\mathbf{V}_m^{\mathsf H}$ and choose $\mathbf{U}_m^{(N_{\mathrm s})}\!\triangleq\![\mathbf{u}_{m,1}\;\cdots\;\mathbf{u}_{m,N_{\mathrm s}}]$ and $\mathbf{V}_m^{(N_{\mathrm s})}\!\triangleq\![\mathbf{v}_{m,1}\;\cdots\;\mathbf{v}_{m,N_{\mathrm s}}]$ as the matrices formed by the $N_{\mathrm s}$ dominant left/right singular vectors. Then, we set $\mathbf{W}_{\mathrm{BB}}[m]=\mathbf{U}_m^{(N_{\mathrm s})}$ and $\mathbf{F}_{\mathrm{BB}}[m]=\mathbf{V}_m^{(N_{\mathrm s})}$ with per-subcarrier power normalization, which yields $\mathbf{H}[m]=\boldsymbol{\Sigma}_m^{(N_{\mathrm s})}$, the principal $N_{\mathrm s}\!\times\!N_{\mathrm s}$ diagonal block of $\boldsymbol{\Sigma}_m$, and white post-combining noise, serving as a bound for the hybrid schemes.

    \item Hybrid \gls{fc}: This hybrid architecture retains frequency–flat, fully connected analog networks $\mathbf{F}_{\mathrm{RF}}\!\in\!\mathbb{C}^{Q_{\mathrm t}Q^2\times Q_{\mathrm t}}$ and $\mathbf{W}_{\mathrm{RF}}\!\in\!\mathbb{C}^{Q_{\mathrm r}Q^2\times Q_{\mathrm r}}$ (constant-modulus). Wideband subspaces are captured via $\mathbf{R}_v=\sum_m \mathbf{V}_m^{(Q_{\mathrm t})}(\mathbf{V}_m^{(Q_{\mathrm t})})^{\mathsf H}$ and $\mathbf{R}_u=\sum_m \mathbf{U}_m^{(Q_{\mathrm r})}(\mathbf{U}_m^{(Q_{\mathrm r})})^{\mathsf H}$ from the SVD $\widecheck{\mathbf{H}}[m]=\mathbf{U}_m\boldsymbol{\Sigma}_m\mathbf{V}_m^{\mathsf H}$; the RF beams are the principal eigendirections of $\mathbf{R}_u,\mathbf{R}_v$ projected onto the unit-modulus manifold. Within these RF spans, per-subcarrier $\mathbf{F}_{\mathrm{BB}}[m],\mathbf{W}_{\mathrm{BB}}[m]$ are obtained by least squares. A single \gls{qrd}-based post-processing step is applied: first, $\mathbf{W}_{\rm RF}\mathbf{W}_{\rm BB}[m]$ is orthonormalized to whiten the post-combining noise, and second, $\mathbf{F}_{\rm RF}\mathbf{F}_{\rm BB}[m]$ is re-scaled to satisfy the per-subcarrier power constraint, ensuring consistency with \eqref{eq:H_detect}.

    \item \gls{fc}+\gls{wmmse}: With frequency–flat, fully connected $\mathbf{F}_{\mathrm{RF}}$ and $\mathbf{W}_{\mathrm{RF}}$ fixed, the baseband weights are refined per subcarrier via alternating \gls{wmmse} under colored noise at the \gls{rf} output. Let $\bar{\mathbf{H}}[m]\!=\!\mathbf{W}_{\mathrm{RF}}^{\mathsf H}\widecheck{\mathbf{H}}[m]\mathbf{F}_{\mathrm{RF}}=\widehat{\mathbf{H}}_{\mathrm{SA}}[m]$ and $\mathbf{G}\!=\!\mathbf{W}_{\mathrm{RF}}^{\mathsf H}\mathbf{W}_{\mathrm{RF}}$. Given $\mathbf{F}_{\mathrm{BB}}[m]$, the receive update uses the \gls{mmse} combiner for $(\bar{\mathbf{H}}[m],\mathbf{G})$; the weight matrix follows from the error covariance; and the transmit update solves the regularized normal equations with a power Lagrange multiplier, yielding $\mathbf{W}_{\mathrm{BB}}[m]$ and $\mathbf{F}_{\mathrm{BB}}[m]$ that minimize the \gls{wmmse} subject to the per–subcarrier power constraint. Each iteration ends with a single QR normalization: $\mathbf{W}_{\mathrm{RF}}\mathbf{W}_{\mathrm{BB}}[m]=\mathbf{Q}_w[m]\mathbf{R}_w[m]$ (whitening) and a rescaling of $\mathbf{F}_{\mathrm{RF}}\mathbf{F}_{\mathrm{BB}}[m]\!\to\!\mathbf{Q}_f[m]$ (power compliance). Each iteration is followed by a single \gls{qrd}-based normalization step, which whitens the post-combining noise and re-scales the effective precoder to meet the per-subcarrier power constraint \cite{OverviewMMW_MIMO,PC_WMMSE}. While the \gls{fc} hybrid offers high beamforming flexibility and near–optimal performance, its dense analog interconnects entail considerable hardware cost and insertion loss.
    
    \item \gls{pc}+\gls{omp}: One \gls{rf} chain drives one \gls{sa}; $\mathbf{F}_{\mathrm{RF}},\mathbf{W}_{\mathrm{RF}}$ are block-diagonal as in \eqref{eq:block_RF}. Analog beams are selected from a sectorized array-response dictionary using a \gls{somp} rule that maximizes the wideband projection matrices $\mathbf{P}_{\mathrm{Tx}}\!=\!\sum_{m}\mathbf{V}_m^{(N_{\mathrm s})}\big(\mathbf{V}_m^{(N_{\mathrm s})}\big)^{\mathsf H}$ and $\mathbf{P}_{\mathrm{Rx}}\!=\!\sum_{m}\mathbf{U}_m^{(N_{\mathrm s})}\big(\mathbf{U}_m^{(N_{\mathrm s})}\big)^{\mathsf H}$, following spatially sparse hybrid precoding \cite{OverviewMMW_MIMO}. With these analog blocks fixed, $\widehat{\mathbf{H}}_{\mathrm{SA}}[m]$ is formed by \eqref{eq:H_SA_def} and detection uses $\mathbf{H}[m]$ in \eqref{eq:H_detect}.

    \item \gls{pc}+\gls{wmmse}: this variant refines only the digital precoder and combiner through classical per-subcarrier \gls{wmmse} alternating updates under colored noise, followed by a single \gls{qrd}-based whitening and normalization step. This procedure corresponds to the standard spectral-efficiency--\gls{wmmse} equivalence specialized for partially-connected hybrid architectures~\cite{PC_WMMSE}.
\end{itemize}
Fig.~\ref{fig:precoding} summarizes \gls{ber} performance for a 16x16 \gls{sa}s 16x16 \gls{um-mimo} \gls{thz} \gls{los} channel. The ordering is consistent and intuitive: \gls{fd} remains the benchmark; among hybrids, \gls{fc}{+}\gls{wmmse} is best, followed by regular \gls{fc}, then \gls{pc}{+}\gls{wmmse}, and finally \gls{pc}{+}\gls{omp}. At \unit{$10^{-3}$}{BER}, \gls{fd} reaches the target near \unit{13}{dB}; \gls{fc}{+}\gls{wmmse} requires about \unit{26}{dB}, while regular \gls{fc} sits roughly \unit{4}{dB} higher for \gls{lord} and around \unit{10}{dB} higher for \gls{zf}. Moving to partially–connected designs, \gls{lord} under \gls{pc}{+}\gls{wmmse} stands at \unit{46}{dB} while \gls{zf} is at around \unit{48}{dB}, and \gls{pc}{+}\gls{omp} is the most power–hungry at \unit{50}{dB} with \gls{lord} and \unit{65}{dB} with \gls{zf}. Two additional takeaways are visible across all hybrids: first, the digital \gls{wmmse} refinement materially improves conditioning—about \unit{4-10}{dB} within \gls{fc} and a much larger \unit{5-15}{dB} within \gls{pc} relative to \gls{omp}—consistent with its role as a spectral-efficiency surrogate under fixed analog networks \cite{PC_WMMSE}; second, when the effective channel is not fully diagonalized, \gls{lord} is markedly more resilient than \gls{zf}, with gains on the order of \unit{7-12}{dB} for \gls{fc} and \unit{6-15}{dB} for \gls{pc} at the same BER. In short, fully–connected analog networks plus a light digital \gls{wmmse} polishing come closest to the \gls{fd} bound, while \gls{pc} designs benefit substantially from \gls{wmmse} but remain limited by \gls{sa} coupling; these impairments increase inter–stream interference and colored noise at the detector input, which is precisely where \gls{lord} outperforms \gls{zf} \cite{OverviewMMW_MIMO,6717211,6847111}.

\subsection{Imperfect CSI and THz channel estimation}

\begin{figure}[t!]
  \centering
    \includegraphics[width=0.48\textwidth]{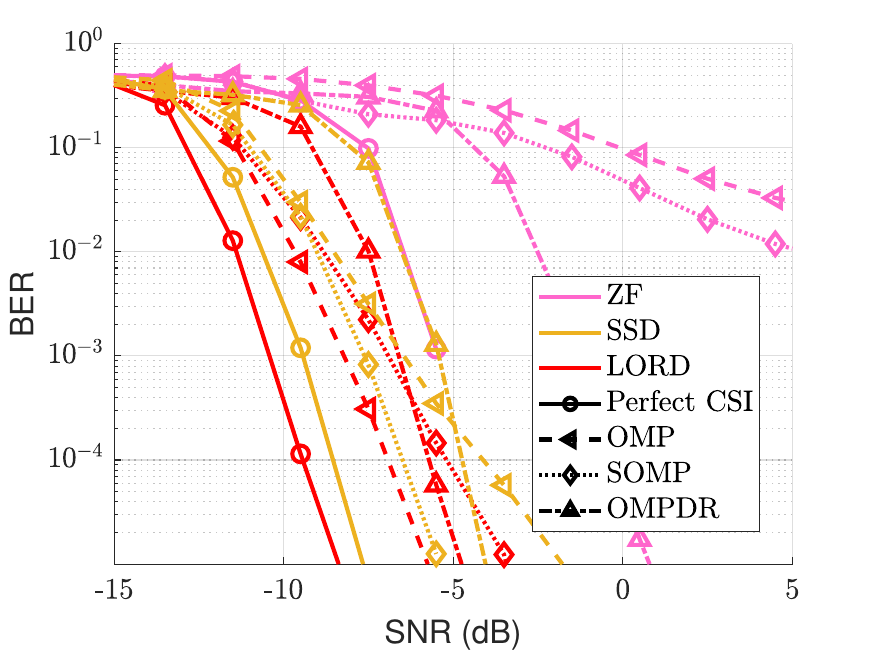}
 \caption{Performance comparison with imperfect CSI: 4x4 \gls{um-mimo}, \gls{thz} \gls{los} channel, $f_{\mathrm{c}}=0.3$ THz, and $D=0.15$ m. }
  \label{fig:imperfect_CSI}
\end{figure}

\gls{um-mimo} channel estimation is particularly challenging under \gls{thz}-band constraints and ultra-large apertures, especially within the \gls{nf} regime~\cite{Tarboush2024Cross}. To evaluate the effect of imperfect \gls{csi} on \gls{mimo} detection, we consider \gls{cs} estimators tailored to the adopted \gls{hspwm} model for \gls{aosas} architectures~\cite{10096832,tarboush9591285}. Three estimators, compatible with the hybrid architecture, are considered in this work: 
\begin{itemize}
    \item The first is the \gls{omp} that operates per subcarrier and iteratively selects steering atoms that best correlate with the pilot observations, refining their complex gains by least squares. Its simplicity and speed come at the cost of ignoring joint structure across subcarriers and \gls{sa}s.
    \item The second estimator is \gls{somp}. It couples all subcarriers during atom selection by maximizing a joint correlation measure, thereby enforcing a common wideband support that aligns with \gls{hspwm} where angles and ranges vary slowly across the band; this stabilizes support detection and reduces \gls{nmse} relative to per-subcarrier \gls{omp} without increasing the number of pilots.
    \item The third estimation technique is \gls{omp-dr} that performs a dictionary-reduction stage: the support is first estimated from a reference Tx/Rx \gls{sa} pair on an oversampled dictionary, and the \gls{aoa}/\gls{aod} bins around the detected support are retained to form a reduced dictionary $\bar{\mathbf{A}}_{\!T,\mathrm{DR}}$ and $\bar{\mathbf{A}}_{\!R,\mathrm{DR}}$ for subsequent tones and \gls{sa}s. This procedure reduces mutual coherence, concentrates training energy, and achieves the lowest \gls{nmse} in our experiments.
\end{itemize}   
The reconstruction of the estimated channel proceeds at the \gls{ae} level and then maps into the forms used by detection. From the pilot data we obtain the \gls{ae}-domain \gls{um-mimo} estimate $\widehat{\widecheck{\mathbf{H}}}[m]$. The post-\gls{rf} \gls{sa}-level channel used throughout the system model is then $\widehat{\mathbf{H}}_{\mathrm{SA}}[m]=\mathbf{W}_{\mathrm{RF}}^\Hpow\,\widehat{\widecheck{\mathbf{H}}}[m]\,\mathbf{F}_{\mathrm{RF}}$, and the baseband detection channel becomes $\widehat{\mathbf{H}}[m]=\mathbf{W}_{\mathrm{BB}}^\Hpow[m]\,\widehat{\mathbf{H}}_{\mathrm{SA}}[m]\,\mathbf{F}_{\mathrm{BB}}[m]$ \eqref{eq:H_SA_def}–\eqref{eq:H_detect}. Fig.~\ref{fig:imperfect_CSI} plots coded \gls{ber} when the received data are generated with the true \gls{sa}-level channel while detection uses the estimated one (\gls{omp}/\gls{somp}/\gls{omp-dr}). Three detectors are compared: linear ZF; LORD, and SSD. Two consistent trends emerge. The first is that the estimator quality dictates the SNR gap. \gls{omp} yields the largest mismatch; \gls{somp} reduces the gap by enforcing joint wideband support; \gls{omp-dr} is closest to perfect CSI for all detectors. Secondly, the detector robustness differs. ZF is mismatch-limited: using $\widehat{\mathbf{H}}$ in $(\widehat{\mathbf{H}}^\mathsf{H}\widehat{\mathbf{H}})^{-1}\widehat{\mathbf{H}}^\mathsf{H}$ introduces residual interference that scales with \gls{nmse} and the condition number of $\widehat{\mathbf{H}}$, producing early error floors with \gls{omp} that recede as we move to \gls{somp} and vanish with \gls{omp-dr}. LORD and SSD maintain steep slopes under the same \gls{csi} because their QR-based metrics and layer ordering mitigate inversion bias; SSD is at least as robust as LORD due to the WRD noise-preservation property. Overall, THz \gls{aosas} links should pair joint-support, dictionary-reduced training (\gls{somp}/\gls{omp-dr}) with non-linear detection (SSD/LORD) to retain near-perfect-\gls{csi} performance at practical pilot budgets, while \gls{zf} requires substantially more accurate \gls{csi} to avoid mismatch-induced errors.

\subsection{MIMO performance analysis: Theoretical and empirical bounds}

\begin{figure}[t!]
  \centering
    \includegraphics[width=0.48\textwidth]{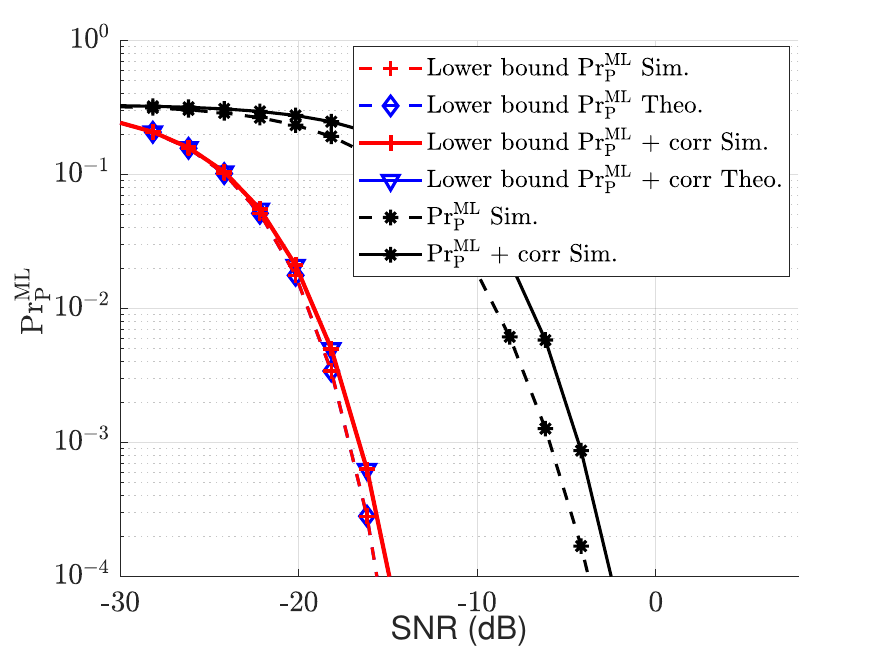}
 \caption{$\mathrm{Pr}_{\mathrm{P}}^{\mathrm{ML}}$ for 16×16 MIMO, THz indoor $\alpha\text{-}\mu$ channel, $F_{\mathrm{c}}=0.142$ THz, and $D=3.15$m}
  \label{fig:ML_Prp}
\end{figure}

We investigate in Fig. \ref{fig:ML_Prp} the pairwise error probability $\mathrm{Pr}_{\mathrm{P}}^{\mathrm{ML}}$ for a 16$\times$16 MIMO \gls{thz} indoor channel following the $\alpha\mhyphen\mu$ small-scale fading under two configurations: the first with independent channel entries, while the second embeds exponential correlation with $\rho_{\mathrm{r}}=\rho_{\mathrm{t}}=0.85$. Three curves are shown for each case: simulated $\mathrm{Pr}_{\mathrm{P}}^{\mathrm{ML}}$, simulated lower bound, and theoretical lower bound. It can be observed that the theoretical lower bound closely matches the simulated lower bound, thereby validating the tightness and accuracy of the proposed analytical approximation. Setting $K=2$ in~\eqref{eq:lower_bound_ML_2} and determining the parameters $c_m$ and $\omega_m$ via moment matching by solving~\eqref{eq:moment_matching_sys} yields a closed-form expression involving the Fox-H function, which provides an excellent match to the simulation results. The results of Fig.~\ref{fig:ML_Prp} further demonstrate that the presence of spatial correlation substantially increases the error probability. This degradation arises because correlation reduces the effective rank of the channel matrix, thereby limiting the spatial diversity that can be exploited by the detector. In highly correlated regimes, the diversity order collapses, and the system becomes more sensitive to fading fluctuations, ultimately resulting in poorer error-rate performance. Nonetheless, as will be shown later, the performance degradation caused by high correlation can be effectively mitigated through the use of puncturing and subspace detection, which enhance parallelizability and yield promising performance gains.

\begin{figure}[t!]
  \centering
    \includegraphics[width=0.48\textwidth]{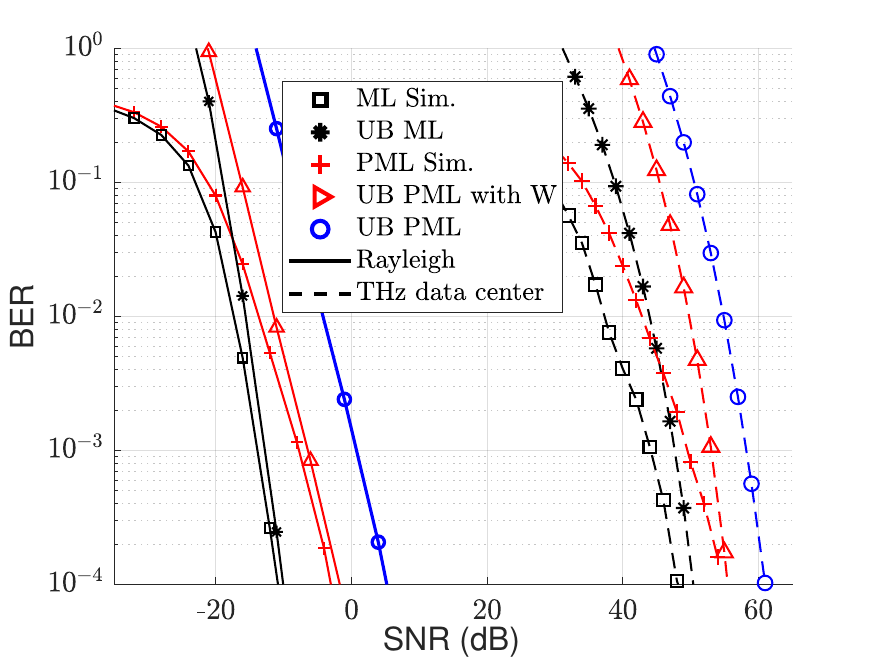}
 \caption{$\mathrm{Pr}_{\mathrm{e}}$ for 4x4 MIMO, Rayleigh vs THz data center channels}
  \label{fig:performance_analysis_MLvsPML_ray_thz}
\end{figure}

Fig. \ref{fig:performance_analysis_MLvsPML_ray_thz} compares the performance of the ML and PML detectors using the empirical bounds from \eqref{eq:ML_empirical} and \eqref{eq:PML_empirical}. Simulated results for ML closely track the derived upper bound, indicating that the $Q\mhyphen$function approximation \cite{1210748} provides a tight estimate of the error probability across a wide range of \gls{snr} values. ML detection outperforms PML, where the V-shaped channel puncturing pattern in the latter, though enhances parallelizability, induces a noticeable performance gap. The tighter upper bound for ML is due to the unitary matrix $\mathbf{Q}$ from QRD, unlike the non-unitary $\mathbf{W}$ in PML, which loosens the bound. Despite performance degradation, PML offers significant complexity reduction, maintaining acceptable \gls{ber} at moderate to high \gls{snr} values. The simulated performance gap between ML and PML is larger in Rayleigh fading (about \unit[8]{dB}) than in \gls{thz} channels (\unit[5]{dB}). In Rayleigh, the gap between \eqref{eq:ML_empirical} and \eqref{eq:PML_empirical} is \unit[17]{dB}, compared to \unit[12]{dB} in \gls{thz}. These results show that performance degradation is less pronounced in \gls{thz} environments. The trade-off between complexity and detection performance is clear, with PML converging more slowly but achieving reasonable \gls{ber} at higher \gls{snr}s. The red-dashed triangular-symbol plot corresponds to evaluating \eqref{eq:Prp_PML} without applying the matrix norm upper bound in step (a), i.e., by directly including the full effect of the matrix $\mathbf{W}$. This tighter treatment significantly improves the accuracy of the approximation, reducing the gap with simulation to less than \unit[0.5]{dB}.

\section{Conclusion}
\label{sec:conclusion}
This work investigates the performance and complexity trade-offs in \gls{thz}-band data detection, accounting for the unique characteristics of \gls{thz} \gls{mimo} channels across different environments. Indoor and outdoor \gls{thz} channels are modeled using the $\alpha\mhyphen\mu$ and \gls{mg} distributions, respectively. The role of spatial parallelizability in reducing overall complexity is emphasized, and a channel-matrix reuse strategy is proposed for \gls{thz} wideband \gls{mimo} systems to reduce computational overhead. Bounds on detection error probability in \gls{thz} \gls{mimo} systems are derived and analyzed. Empirical simulations under practical \gls{thz} conditions demonstrate that channel-matrix puncturing in \gls{ssd} improves performance in non-orthogonal \gls{los} and data center scenarios while incurring minimal performance loss in ill-conditioned \gls{thz} channels. These performance gains are achieved alongside significant complexity reduction. Furthermore, the complexity analysis of candidate data detectors highlights the critical trade-offs between performance, complexity, and latency for efficient baseband processing, towards enabling \gls{tbps} data rates in \gls{thz} communication systems. In addition, we quantified robustness to imperfect \gls{csi} and practical hybrid precoding/combining: LORD/SSD remained reliable under \gls{omp}/\gls{somp}/\gls{omp-dr} estimates, while ZF was more sensitive; partially connected chains with \gls{wmmse}/\gls{omp} approached the \gls{fd}/\gls{fc} references within hardware limits. Future work may explore incorporating MC-aware architectures and assessing their influence on estimation, hybrid design, and detection performance.

\section*{Acknowledgment}
The authors would like to thank Prof. Christoph Studer from ETH Zurich for his valuable feedback and insightful comments, which helped improve the quality of this work.

\bibliographystyle{IEEEtran}
\bibliography{IEEEabrv,my_bibliography}

\begin{thebibliography}{10}
\providecommand{\url}[1]{#1}
\csname url@samestyle\endcsname
\providecommand{\newblock}{\relax}
\providecommand{\bibinfo}[2]{#2}
\providecommand{\BIBentrySTDinterwordspacing}{\spaceskip=0pt\relax}
\providecommand{\BIBentryALTinterwordstretchfactor}{4}
\providecommand{\BIBentryALTinterwordspacing}{\spaceskip=\fontdimen2\font plus
\BIBentryALTinterwordstretchfactor\fontdimen3\font minus \fontdimen4\font\relax}
\providecommand{\BIBforeignlanguage}[2]{{%
\expandafter\ifx\csname l@#1\endcsname\relax
\typeout{** WARNING: IEEEtran.bst: No hyphenation pattern has been}%
\typeout{** loaded for the language `#1'. Using the pattern for}%
\typeout{** the default language instead.}%
\else
\language=\csname l@#1\endcsname
\fi
#2}}
\providecommand{\BIBdecl}{\relax}
\BIBdecl

\bibitem{JemaaDetection2022}
H.~Jemaa~\emph{et al.}, ``{THz}-band, {T}bps {MIMO} communications: a joint data detection and decoding framework,'' in \emph{Proc. Asilomar Conf. Signals, Systems and Computers (Asilomar)}, 2022, pp. 665--669.

\bibitem{akyildiz2022terahertz}
I.~F. Akyildiz~\emph{et al.}, ``Terahertz band communication: an old problem revisited and research directions for the next decade,'' \emph{IEEE Trans. Commun.}, vol.~70, no.~6, pp. 4250--4285, 2022.

\bibitem{sarieddeen2020overview}
H.~Sarieddeen, M.-S. Alouini, and T.~Y. Al-Naffouri, ``{A}n overview of signal orocessing techniques for terahertz communications,'' \emph{Proc. IEEE}, vol. 109, no.~10, pp. 1628--1665, 2021.

\bibitem{Jornet2024Evolution}
J.~M. Jornet~\emph{et al.}, ``{T}he evolution of applications, hardware design, and channel modeling for terahertz ({TH}z) band communications and sensing: ready for 6{G}?'' \emph{Proc. IEEE}, pp. 1--32, 2024.

\bibitem{rajatheva2020white}
N.~Rajatheva~\emph{et al.}, ``White paper on broadband connectivity in {6G},'' \emph{arXiv preprint arXiv:2004.14247}, 2020.

\bibitem{sarieddeen2023bridging2}
H.~Sarieddeen~\emph{et al.}, ``{B}ridging the complexity gap in {Tbps-achieving THz-band} baseband processing,'' \emph{IEEE Wireless Commun. Mag.}, vol.~31, no.~5, pp. 287--294, 2024.

\bibitem{tarboush9591285}
S.~Tarboush~\emph{et al.}, ``{T}era{MIMO}: A channel simulator for wideband ultra-massive {MIMO} terahertz communications,'' \emph{IEEE Trans. on Vehic. Technol.}, vol.~70, no.~12, pp. 12\,325--12\,341, 2021.

\bibitem{sheikh2022thz}
F.~Sheikh~\emph{et al.}, ``{THz} measurements, antennas, and simulations: from the past to the future,'' \emph{IEEE Journal of Microwaves}, vol.~3, no.~1, pp. 289--304, 2022.

\bibitem{Jemaa2024Performance}
H.~Jemaa~\emph{et al.}, ``{P}erformance analysis of outdoor {THz} links under mixture {G}amma fading with misalignment,'' \emph{{IEEE} Commun. Lett.}, vol.~28, no.~11, pp. 2668--2672, 2024.

\bibitem{papasotiriou2023outdoor}
E.~N. Papasotiriou~\emph{et al.}, ``Outdoor {THz} fading modeling by means of {G}aussian and {G}amma mixture distributions,'' \emph{Sci. Rep.}, vol.~13, no.~1, p. 6385, 2023.

\bibitem{papasotiriou2021experimentally}
------, ``An experimentally validated fading model for {THz} wireless systems,'' \emph{Sci. Rep.}, vol.~11, no.~1, p. 18717, 2021.

\bibitem{Lu10496996}
H.~Lu~\emph{et al.}, ``A tutorial on near-field {XL-MIMO} communications toward {6G},'' \emph{IEEE Commun. Surveys Tuts.}, vol.~26, no.~4, pp. 2213--2257, 2024.

\bibitem{9763525}
Z.~Dong and Y.~Zeng, ``{N}ear-field spatial correlation for extremely large-scale array communications,'' \emph{IEEE Commun. Lett.}, vol.~26, no.~7, pp. 1534--1538, 2022.

\bibitem{tarboush2022single}
S.~Tarboush~\emph{et al.}, ``{S}ingle- versus multicarrier terahertz-band communications: a comparative study,'' \emph{{IEEE} Open J. of the Commun. Soc.}, vol.~3, pp. 1466--1486, 2022.

\bibitem{Wang2022HybridBeamforming}
P.-H. Chang and T.-D. Chiueh, ``{H}ybrid beamforming for wideband terahertz massive {MIMO} communications with low-resolution ohase shifters and true-time-delay,'' \emph{IEEE Trans. Wirel. Commun.}, vol.~23, no.~7, pp. 8000--8012, 2024.

\bibitem{Han2016Multi}
C.~Han, A.~O. Bicen, and I.~F. Akyildiz, ``{M}ulti-wideband waveform design for distance-adaptive wireless communications in the terahertz band,'' \emph{IEEE Trans. Signal Process.}, vol.~64, no.~4, pp. 910--922, 2016.

\bibitem{7244171}
S.~Yang and L.~Hanzo, ``{F}ifty years of {MIMO} detection: the road to large-scale {MIMO}s,'' \emph{IEEE Commun. Surveys Tuts.}, vol.~17, no.~4, pp. 1941--1988, 2015.

\bibitem{7472341}
Y.~Cao, W.~Su, and S.~N. Batalama, ``{D}istributed {MIMO} systems: receiver design and {ML} detection,'' in \emph{Proc. IEEE Int. Conf. Acoustics, Speech, and Signal Process. (ICASSP)}, 2016, pp. 3566--3570.

\bibitem{Sarieddeen8765243}
H.~Sarieddeen, M.-S. Alouini, and T.~Y. Al-Naffouri, ``{T}erahertz-band ultra-massive spatial modulation {MIMO},'' \emph{{IEEE} J. Sel. Areas Commun.}, vol.~37, no.~9, pp. 2040--2052, 2019.

\bibitem{jeon2015optimality}
C.~Jeon~\emph{et al.}, ``{O}ptimality of large {MIMO} detection via approximate message passing,'' in \emph{Proc. IEEE Int. Symp. on Inf. Theory (ISIT)}, 2015, pp. 1227--1231.

\bibitem{7755889}
M.~Wu~\emph{et al.}, ``{H}igh-throughput data detection for massive {MU}-{MIMO}-{OFDM} using coordinate descent,'' \emph{{IEEE} Trans. Circuits Syst. {I}}, vol.~63, no.~12, pp. 2357--2367, 2016.

\bibitem{9148630}
L.~V. Nguyen, D.~H.~N. Nguyen, and A.~L. Swindlehurst, ``{SVM}-based channel estimation and data detection for massive {MIMO} systems with one-bit {ADCs},'' in \emph{Proc. IEEE Int. Conf. Commun. (ICC)}, 2020, pp. 1--6.

\bibitem{kobayashi2016}
R.~T. Kobayashi and T.~Abr{\~a}o, ``Ordered {MMSE}--{SIC} via sorted {QR} decomposition in ill conditioned large-scale {MIMO} channels,'' \emph{Telecommunication systems}, vol.~63, pp. 335--346, 2016.

\bibitem{hu2017}
S.~Hu and F.~Rusek, ``{A} soft-output {MIMO} detector with achievable information rate based oartial marginalization,'' \emph{IEEE Trans. Signal Process.}, vol.~65, no.~6, pp. 1622--1637, 2017.

\bibitem{ammari2015analysis}
M.~L. Ammari and P.~Fortier, ``On the analysis of {MIMO}-{ZF} receiver over fully correlated {MIMO} {R}ayleigh fading with {LMMSE} channel estimation,'' \emph{Wirel. Pers. Commun.}, vol.~85, pp. 1025--1042, 2015.

\bibitem{9896734}
Y.~Wu~\emph{et al.}, ``{3-D} hybrid beamforming for terahertz broadband communication system with beam squint,'' \emph{IEEE Trans. Broadcast.}, vol.~69, no.~1, pp. 264--275, 2023.

\bibitem{Sarieddeen8186206}
H.~Sarieddeen, M.~M. Mansour, and A.~Chehab, ``{L}arge {MIMO} detection schemes based on channel ouncturing: oerformance and complexity analysis,'' \emph{IEEE Trans. Commun.}, vol.~66, no.~6, pp. 2421--2436, 2018.

\bibitem{prudnikov1986integrals}
{Prudnikov \emph{et al.}, Anatoli{\u\i} Platonovich }, \emph{Integrals and series: special functions}.\hskip 1em plus 0.5em minus 0.4em\relax CRC press, 1986, vol.~2.

\bibitem{kilbas2004h}
A.~A. Kilbas, \emph{H-transforms: Theory and Applications}.\hskip 1em plus 0.5em minus 0.4em\relax CRC press, 2004.

\bibitem{5090420}
A.~M. Magableh and M.~M. Matalgah, ``Moment generating function of the generalized $\alpha$--$\mu$ distribution with applications,'' \emph{IEEE Commun. Lett.}, vol.~13, no.~6, pp. 411--413, 2009.

\bibitem{Tarboush2024Cross}
S.~Tarboush, A.~Ali, and T.~Y. Al-Naffouri, ``Cross-field channel estimation for ultra {massive-MIMO THz} systems,'' \emph{IEEE Trans. Wirel. Commun.}, vol.~23, no.~8, pp. 8619--8635, 2024.

\bibitem{4787140}
Y.~H. Gan, C.~Ling, and W.~H. Mow, ``{C}omplex lattice reduction algorithm for low-complexity full-diversity {MIMO} detection,'' \emph{IEEE Trans. Signal Process.}, vol.~57, no.~7, pp. 2701--2710, 2009.

\bibitem{1603705}
Z.~Guo and P.~Nilsson, ``Algorithm and implementation of the {K}-best sphere decoding for {MIMO} detection,'' \emph{{IEEE} J. Sel. Areas Commun.}, vol.~24, no.~3, pp. 491--503, 2006.

\bibitem{shen2007generalized}
C.~Shen, M.~P. Fitz, and M.~Siti, ``{G}eneralized soft-output layered orthogonal lattice detector for golden code,'' in \emph{Proc. IEEE Wireless Commun. and Netw. Conf. (WCNC)}, 2007, pp. 525--529.

\bibitem{izadinasab2019partial}
M.~K. Izadinasab and O.~Damen, ``{P}artial lattice reduction and subspace detection of large-scale {MIMO} systems,'' in \emph{Proc. IEEE Int. Symp. Personal Indoor and Mobile Radio Commun. (PIMRC)}, 2019, pp. 1--6.

\bibitem{wolniansky1998v}
P.~W. Wolniansky~\emph{et al.}, ``{V}-{B}{L}{A}{S}{T}: an architecture for realizing very high data rates over the rich-scattering wireless channel,'' in \emph{Proc. of ISSSE}, 1998, pp. 295--300.

\bibitem{mansour2020low}
M.~M. Mansour, ``low-complexity soft-output {MIMO} detectors based on optimal channel ouncturing,'' \emph{IEEE Trans. Wirel. Commun.}, vol.~20, no.~4, pp. 2729--2745, 2020.

\bibitem{1210748}
M.~Chiani, D.~Dardari, and M.~K. Simon, ``{N}ew exponential bounds and approximations for the computation of error probability in fading channels,'' \emph{IEEE Trans. Wirel. Commun.}, vol.~2, no.~4, pp. 840--845, 2003.

\bibitem{payami2020accurate}
M.~Payami and A.~Falahati, ``Accurate variable-order approximations to the sum of $\alpha$--$\mu$ variates with application to {MIMO} systems,'' \emph{IEEE Trans. Wirel. Commun.}, vol.~20, no.~3, pp. 1612--1623, 2020.

\bibitem{website_wolfram}
\BIBentryALTinterwordspacing
Meijer {G}-function: integration. [Online]. Available: \url{https://functions.wolfram.com/07.34.21.0012.01}
\BIBentrySTDinterwordspacing

\bibitem{4518202}
J.~Jald{\'e}n, D.~Seethaler, and G.~Matz, ``Worst- and average-case complexity of {LLL} lattice reduction in {MIMO} wireless systems,'' in \emph{Proc. IEEE Int. Conf. Acoustics, Speech, and Signal Process. (ICASSP)}, 2008, pp. 2685--2688.

\bibitem{douillard2019next}
C.~Douillard, ``{N}ext-generation channel coding towards terabit/s wireless communications,'' in \emph{{GDR} ISIS Workshop: Enabling Technologies for sub-TeraHertz and TeraHertz communications}, 2019.

\bibitem{OverviewMMW_MIMO}
R.~W.~J. Heath, N.~González-Prelcic, S.~Rangan, W.~Roh, and A.~M. Sayeed, ``An overview of signal processing techniques for millimeter wave {MIMO} systems,'' \emph{{IEEE} J. Sel. Topics Signal Process.}, vol.~10, no.~3, pp. 436--453, 2016.

\bibitem{PC_WMMSE}
Z.~Zhao, Z.~Lin, P.~Zhu, and Q.~Zhang, ``Partially-connected hybrid beamforming for spectral efficiency maximization via a weighted {MMSE} equivalence,'' \emph{IEEE Trans. Wirel. Commun.}, vol.~20, no.~8, pp. 5031--5044, 2021.

\bibitem{6717211}
O.~E. Ayach~\emph{et al.}, ``Spatially sparse precoding in millimeter wave mimo systems,'' \emph{IEEE Trans. Wirel. Commun.}, vol.~13, no.~3, pp. 1499--1513, 2014.

\bibitem{6847111}
A.~Alkhateeb~\emph{et al.}, ``Channel estimation and hybrid precoding for millimeter wave cellular systems,'' \emph{{IEEE} J. Sel. Topics Signal Process.}, vol.~8, no.~5, pp. 831--846, 2014.

\bibitem{10096832}
S.~Tarboush, A.~Ali, and T.~Y. Al-Naffouri, ``Compressive estimation of near field channels for ultra massive-mimo wideband thz systems,'' in \emph{Proc. IEEE Int. Conf. Acoustics, Speech, and Signal Process. (ICASSP)}, 2023, pp. 1--5.

\end{thebibliography}

\end{document}